\documentclass[a4paper, amsfonts, amssymb, amsmath, reprint, showkeys, nofootinbib, twoside,superscriptaddress,aip,citeautoscript]{revtex4-1}
\usepackage[english]{babel}
\usepackage[utf8]{inputenc}

\usepackage[colorinlistoftodos, color=green!40, prependcaption]{todonotes}
\usepackage{amsthm}
\usepackage{mathtools}
\usepackage{physics}
\usepackage{xcolor}
\usepackage{graphicx}
\usepackage[left=16mm,right=16mm,top=20mm,bottom=22mm,columnsep=15pt]{geometry} 
\usepackage{adjustbox}
\usepackage{placeins}
\usepackage[T1]{fontenc}
\usepackage{lipsum}
\usepackage{csquotes}

\usepackage[pdftex, pdftitle={Article}, pdfauthor={Author}]{hyperref} 
\usepackage{float}
\newcommand{\kB}{k_{\mathrm{B}}}
 
\setcitestyle{super}
\def\kbt{k_{\mathrm{B}}T}

\newcommand{\vc}[1]{\mathbf{#1}}

\newcommand{\revision}{\textcolor{black} } 

\begin{document}
\title{Variational deep learning of equilibrium transition path ensembles}
\author{Aditya N. Singh}
\affiliation{Department of Chemistry, University of California, Berkeley, CA 94720, USA \looseness=-1}
\affiliation{Chemical Sciences Division, Lawrence Berkeley National Laboratory, Berkeley, CA 94720, USA \looseness=-1}

\author{David T. Limmer}
\email[Electronic mail: ]{dlimmer@berkeley.edu}
\affiliation{Department of Chemistry, University of California, Berkeley, CA 94720, USA \looseness=-1}
\affiliation{Chemical Sciences Division, Lawrence Berkeley National Laboratory, Berkeley, CA 94720, USA \looseness=-1}
\affiliation{Materials Science Division, Lawrence Berkeley National Laboratory, Berkeley, CA 94720, USA \looseness=-1}
\affiliation{Kavli Energy Nanoscience Institute at Berkeley, Berkeley, CA 94720, USA \looseness=-1}

\date{\today} 

\begin{abstract}
We present a time dependent variational method to learn the mechanisms of equilibrium reactive processes and efficiently evaluate their rates within a transition path ensemble. This approach builds off variational path sampling methodology by approximating the time dependent commitment probability within a neural network ansatz. The reaction mechanisms inferred through this approach are elucidated by a novel decomposition of the rate in terms of the components of a stochastic path action conditioned on a transition. This decomposition affords an ability to resolve the typical contribution of each reactive mode and their couplings to the rare event. The associated rate evaluation is variational and systematically improvable through the development of a cumulant expansion. We demonstrate this method in both over- and under-damped stochastic equations of motion, in low-dimensional model systems and the isomerization of solvated alanine dipeptide. In all examples, we find that we can obtain quantitatively accurate estimates of the rates of the reactive events with minimal trajectory statistics, and gain unique insight into the transitions through the analysis of their commitment probability.
\end{abstract}
\maketitle

\section*{Introduction}
In complex systems, understanding the mechanism of transitions between long-lived metastable states is hampered by the general collective nature of the dynamics and the difficulty of observing these rare but important events.\cite{peters2017reaction} While methods like transition path sampling\cite{bolhuis2002transition} exist to harvest rare events computationally, their distillation into mechanistic descriptions is cumbersome, and the conversion of that description into quantitative statements of their rate is challenging.\cite{geissler1999kinetic,vanden2010transition}  Here, we present a method that uses a neural-network ansatz with a variational optimization procedure to compute the time dependent commitment probability from a reactive trajectory ensemble. The method involves learning a unique policy, in the form of an optimal external control force, that reweights a reactive conditioned path ensemble to an unconditioned ensemble that reacts autonomously. The optimal force is simply related to the commitment probability,\cite{majumdar2015effective,das2022direct} and serves as an ideal descriptor of the reaction. The reweighting principle developed within the framework of variational path sampling\cite{das2019variational} is expressed in terms of the stochastic action, which allows us to decompose the rate into additive contributions from different degrees of freedom, including collective coordinates that describe molecular transitions. This decomposition provides a means of identifying relevant order parameters without making \textit{a-priori} assumptions. The combination of the mechanistic insight afforded by an interpretable representation of the reaction and the validation through a variational evaluation of the rate, provides a robust method for distilling features of equilibrium transition path ensembles. 

The investigation of reactive events requires access to timescales that are considerably longer than the local relaxation time of the system. The canonical approach to investigate these processes has leveraged physically intuitive low-rank descriptions of the system to infer mechanistic insight, and bridge the timescales through reactive flux calculations or importance sampling.\cite{eyring1935activated,chandler1978statistical,hanggi1990reaction,carter1989constrained,zwanzig2001nonequilibrium} The notion of an ideal reaction coordinate capable of providing a complete description of the reactive event dates back to Onsager\cite{onsager1938initial}, and was formalized within the context of chemical physics as the committor-- a map between the phase space position of a system and the likelihood of it reacting\cite{geissler1999kinetic,bolhuis2000reaction,geissler2001autoionization,vanden2006towards}. Learning this high dimensional function has attracted interest from a diversity of fields, and significant \revision{advances} has been made through methods that employ importance sampling and machine learning.\cite{weinan2002string,ma2005automatic,vanden2006towards,coifman2008diffusion,wales2009calculating,jung2019artificial,khoo2019solving,li2019computing,rotskoff2022active,hasyim2022supervised,vani2022computing,falkner2022conditioning,evans2022computing,li2019computing,li2022semigroup,strahan2023predicting,thiede2019galerkin,chen2023committor,strahan2021long,jung2023machine,strahan2023inexact}Some notable approaches have leveraged the confinement of the transition region to compute it using string methods\cite{weinan2002string,vanden2006towards,vanden2006transition}, coarse-grained the phase-space to approximate it through diffusion maps\cite{coifman2008diffusion,evans2022computing,trstanova2020local,lai2018point}, and parameterized neural-networks by either fitting the committor directly\cite{ma2005automatic,jung2019artificial,jung2023machine} or solving the variational form of the steady-state backward Kolmogorov equation\cite{khoo2019solving} by combining it with importance sampling methods\cite{li2019computing,rotskoff2022active,hasyim2022supervised}. While the learning procedures applied previously have been successful in fitting high dimensional representations of the reaction coordinate or committors, their nonlinearity has largely resulted in a difficulty in interpreting the relative importance of physically distinct descriptors and converting those descriptors into a robust measure of the rate. Earlier developments of methods based on likelihood maximization \cite{ma2005automatic,peters2006obtaining,peters2007extensions} have offered linear ways to make this analysis tractable to complex processes\cite{jungblut2013optimising, leitold2014folding, lechner2010nonlinear,peters2016reaction, bolhuis2015practical}.  However, these approaches have overwhelmingly relied on physical intuition to express likelihood functions.\cite{peters2016reaction} 

The method that we present builds off of variational path sampling\cite{das2019variational,das2021reinforcement,das2022direct,nemoto2016population, jacobson2019direct} that has provided an alternative approach for sampling rare events. \revision{These methods and related ones employ ideas from stochastic optimal control\cite{holdijk2022path,yan2022learning,zhong2022limited,vargas2021solving}, and are most useful in investigating nonequilibrium steady states as they do not invoke detailed balance.} Of particular interest is a recent method\cite{das2022direct} that has detailed how to express a low-rank ansatz for an optimal control force to drive rare events and estimate their rates. Our work exploits the fact that the optimization of this control force, or policy, is related to the time dependent committor. We find that in equilibrium systems, where path sampling methods afford a way to generate a reference reactive trajectory ensemble, the optimization of this committor becomes straightforward, and allows the use of a neural-network (NN) ansatz to solve the time-dependent backward Kolmogorov equation\cite{van1992stochastic}, providing a time dependent and probabilistic representation of the reaction. While the method computes a nonlinear function, the form of the optimized loss is given by the difference in stochastic actions that quantifies the distance between a conditioned and a reference trajectory ensemble. 
For systems in which we saturate the variational bound, this quantity is unique and linearly 
decomposable on a per-coordinate basis, and can be understood as a measure of the importance of each coordinate to conditioning a trajectory to be reactive. This metric is purely based on the intrinsic mechanism of the reaction, and can be extended to collective coordinates, allowing us to identify the relevant reaction descriptors without making \textit{a-priori} assumptions.

This paper is organized as follows. First, we review the variational path sampling formalism to discuss the theory behind this method. Next, we validate this method by applying it to a couple of low dimensional systems where numerically exact results are possible. We probe the sensitivity of this method to limited statistics as well as the applicability to systems integrated with underdamped equations of motion. Then, we illustrate how the per-coordinate stochastic action encodes the relevance of a coordinate to the reaction. Finally, we apply this method to study the isomerization of alanine dipeptide in implicit and explicit solvent. In both of these cases, we show how the method can be used to infer a mechanistic picture of the reaction, and identify important reaction descriptors among a redundant set of internal coordinates.

\section{Variational path sampling formalism}
For simplicity, we consider a system evolving under an overdamped Langevin equation of the form,
\begin{equation}\label{eq:1.1}
    \gamma_i  \dot{\vc{r}}_i(t) = \vc{F}_i\left (\mathbf{r}^N \right ) + \boldsymbol{\eta}_i(t)
\end{equation}
where $\dot{\vc{r}}_i$ is the rate of change of the $i$th particle's position at time $t$ in $d$ dimensions, $\gamma_i$ is the friction coefficient, and $\boldsymbol{\eta}_i(t)$ denotes a Gaussian random force with mean $\langle \boldsymbol{\eta}_i (t) \rangle=0$ and variance  \revision{$\langle \boldsymbol{\eta}_i(t)\otimes \boldsymbol{\eta}_j (t') \rangle = 2\gamma_i \kbt \delta_{ij} \vc{1_d} \delta(t-t')$
where $\otimes$ denotes the cross-product operator, $\vc{1_d}$ is an identity-matrix of size $d\times d$ and
$\kbt$ is Boltzmann's constant times the temperature.} The conservative force $\vc{F}_i\left (\mathbf{r}^N \right ) = -\nabla_i V\left (\mathbf{r}^N \right )$ is  given by the gradient of the potential $V\left (\mathbf{r}^N \right )$ with $\vc{r}^N$ the full $N$-particle configuration. We are interested in investigating reactive events, so we consider potentials that exhibit metastability. 

We consider transitions between two metastable states, $A$ and $B$, which in general are collections of configurations defined through the indicator functions $h_A[\mathbf{r}^N(t)]$ and $h_B[\mathbf{r}^N(t)]$, where
\begin{equation}
h_X[\mathbf{r}^N(t)] = \begin{cases} 1 \quad \mathbf{r}^N(t)\in X\\
0 \quad \mathbf{r}^N(t)\notin X \end{cases}
\end{equation}
for $X=\{ A,B\}$. \revision{For the rest of the the paper, indicator functions are going to written down simply as functions of time in favor of brevity.} The rate for the $A \rightarrow B$ transition can be defined by the time derivative of the side-side correlation function,\cite{chandler1978statistical}
\begin{equation}\label{eq:1.2}
k = \frac{d}{dt}\frac{\langle h_A(0) h_B(t)\rangle}{\langle h_A \rangle} =\frac{d}{dt} \langle h_{B|A}(t)\rangle
\end{equation}
where $\langle \cdots \rangle$ denotes an average computed over a stationary distribution and $h_{B|A}$ is the conditional probability of starting in $A$ and ending in $B$ at $t$. Provided a separation of timescales between the local relaxation time within a state, $\tau_{\mathrm{mol}}$, and $1/k$, the rate is given by the path integral
\begin{align}\label{eq:1.3}
    k t_f = \int \mathcal{D}[\mathbf{X}] h_{B|A}(t_f) P [\mathbf{X}]
\end{align}
where when $t_f$ is in the range $\tau_\textrm{mol} < t_f \ll 1/k$, the probability to transition grows linearly with time. The path integral sums over all trajectories $\mathbf{X}=\{\vc{r}^N(0),\dots , \vc{r}^N(t_f)\}$, or the timeseries of the state of the system evolved for time $t_f$, weighted by the likelihood of  observing a trajectory $P [\mathbf{X}]$. This path integral is a trajectory partition function associated with reactive paths,\cite{dellago1999calculation} and equal to the transition probability between $A$ to $B$ in time $t_f$. 

Variational path sampling uses the path partition function representation of the rate together with a dynamical reweighting approach\cite{gao2019nonlinear} to extract reactive paths effectively,\cite{das2021reinforcement} evaluate rates accurately,\cite{das2022direct} and we show here, provide detailed mechanistic information concerning the rare event. Variational path sampling does this by considering the system as before, but under the action of an additional time-dependent drift $\boldsymbol{\lambda}_i(\mathbf{r}^N,t)$, which enters the equation of motion as
\begin{equation}\label{eq:1.4}
    \gamma_i  \dot{\vc{r}}_i = \vc{F}_i(\mathbf{r}^N) + \boldsymbol{\lambda}_i(\mathbf{r}^N,t) + \boldsymbol{\eta}_i(t)
\end{equation}
where the conservative force, noise and friction are the same as the reference system without $\boldsymbol{\lambda}_i(\mathbf{r}^N,t)$. 
For this driven system, the rate $k_\lambda$ between the same two metastable states $A$ and $B$ is given by an analogous relation as in the reference system
\begin{align}\label{eq:1.5}
    k_\lambda t_f = \int \mathcal{D}[\mathbf{X}] h_{B|A}(t) P_\lambda [\mathbf{X}]
\end{align}
where $P_\lambda[\mathbf{X}]$ denotes the probability of observing a trajectory $\mathbf{X}$ integrated using Eq. \ref{eq:1.4}. By virtue of the Girsanov transformation, these two rate expressions can be related to each other. Specifically, using the Radon-Nikodym derivative to define the change in stochastic action, \revision{$\Delta U_\lambda[\mathbf{X}] = \ln P_\lambda[\mathbf{X}]/P[\mathbf{X}]$}, the rate in the driven system can be rewritten as\cite{kuznets2021dissipation}
\begin{align}\label{eq:1.6}
    \ln k_\lambda t_f &= \ln \int D[\mathbf{X}] P[\mathbf{X}] h_{B|A} (t_f) e^{\Delta U_\lambda} \notag \\
    &= \ln k t_f  + \ln\left  \langle e^{\Delta U_\lambda}\right  \rangle_{B|A}
\end{align}
where we have employed \revision{$\langle \dots \rangle_{B|A}=\langle h_{B}(t_f) h_{A}(0) \dots \rangle/\langle h_{A}(0) \rangle$} as a conditional average over a reference reactive ensemble to relate the two rates. For the case of the overdamped Langevin equation, the change in stochastic action is given by a difference of Onsager-Machlup actions\cite{onsager1953fluctuations}

\begin{align}\label{eq:1.7}
    \revision{\Delta U_\lambda[\mathbf{X}] = -\sum_{i=1} ^N \frac{1}{4\gamma_i k_{\mathrm{B}}T} \int_0^{t_f} dt \, |\boldsymbol{\lambda}_i|^2- 2\boldsymbol{\lambda}_i \cdot (\mathbf{\gamma_i \dot r_i} - \mathbf{F_i})}
\end{align}
\revision{where $|\dots|$ denotes the norm of the vector, and $\cdot$ denotes the dot product operator. The discretized form of the OM action can be simplified further by noting that the difference between the the time derivative of the positions and the conservative force is simply given by the noises.}

The relationship between rates in Eq.~\ref{eq:1.6} is exact for any time-dependent drift $\boldsymbol{\lambda}_i(\mathbf{r}^N,t)$. It is distinct from that employed previously,\cite{das2022direct} which related the reference and driven rates to an expectation value in the driven system. In variational path sampling, we consider a class of $\boldsymbol{\lambda}_i(\mathbf{r}^N,t)$ which enforce the transition to occur with probability 1. In such a case, provided access to a reactive path ensemble in which to evaluate the expectation values, the rate in the reference system can be obtained directly as an exponential average, 
 \begin{align}
    \ln k t_f &=-\ln \left \langle e^{\Delta U_\lambda} \right \rangle_{B|A} \label{eq:2.2}\\
    & = -\langle \Delta U_\lambda \rangle_{B|A} - \sum_{n=2}^\infty \frac{1}{n!} \mathcal{C}_{B|A}^{(n)}(\Delta U_\lambda) \label{eq:2.3} 
\end{align}
or a cumulant expansion, where $\mathcal{C}_{B|A}^{(n)}(\Delta U_\lambda)$ denotes the $n$'th cumulant of $\Delta U_\lambda$ averaged in the reactive ensemble. We will refer to these different estimators as $ k^{(\exp)}$ for the exponential average and $k^{(n)}$ for the cumulant expansion where $n$ will denote where the sum was truncated. 

Truncation of the cumulant expansion for $n=1$ provides a variational bound of the rate. This is seen by applying Jensen's inequality to Eq.~\ref{eq:1.6}, 
\begin{align}\label{eq:1.9}
    \revision{\ln kt_f} & \revision{ \leq \ln k_\lambda t_f - \langle \Delta U_\lambda \rangle_{B|A} } \\ \notag
    \revision{\implies \ln kt_f} & \revision{\leq  - \langle \Delta U_\lambda \rangle_{B|A} }
\end{align}
\revision{where we have used conservation of probability $\ln k_{\lambda} t_f \leq 0$ in the second step to eliminate the rate of the driven process from the inequality.} Hence, the rate in the reference system is just the Kullback-Leibler (KL) divergence between the driven and reference path ensembles, or equivalently the mean change in action. In equilibrium systems, this relation is similar to the variational structure of transition state theory, which also provides an upper bound to the rate.\cite{chandler1978statistical} However, this expression is also closely related to the reversible work theorem in equilibrium thermodynamics,\cite{chandler1987introduction}  as it relates the smallest change required to transform one ensemble to another.\cite{jarzynski1997nonequilibrium,crooks1999entropy} In this case, the transformation is between an unconditioned path ensemble and a reactive path ensemble.  Just as the minimum amount of work done on a physical system is given by its reversible limit, which reflects the way in which a system would naturally transform, so too we find the minimum driving force to ensure a reaction is related to the way in which a system would naturally react.\cite{turner2014meta} This is shown by noting that the force that saturates this bound in Eq. ~\ref{eq:1.9} is the Doob force, denoted  $\boldsymbol{\lambda}^*(\mathbf{r}^N,t)$, and is related to the solution of the backward Kolmogorov equation \cite{jack2010large,chetrite2015variational, majumdar2015effective, das2022direct}. For an overdamped Langevin dynamics this is,
\begin{equation}\label{Eq:OBKE}
    \partial_t q(\mathbf{r}^N,t) = - \sum_{i=1}^N \frac{\vc{F}_i\left (\mathbf{r}^N \right )}{\gamma_i} \cdot \nabla_i q(\mathbf{r}^N,t)-  \frac{\kbt}{\gamma_i} \nabla^2_i q(\mathbf{r}^N,t) 
\end{equation}
with boundary conditions $q(\mathbf{r}^N,t_f) = h_B(t_f)$ and $q(\mathbf{r}^N,0) = h_A(0)$. The function that solves this expression, $q(\mathbf{r}^N,t)$, is the time-dependent committor function,\cite{majumdar2015effective,das2022direct} or the probability of reaching state $B$ at $t_f$ given a position $\vc{r}^N$ at time $t$. In the stationary limit, where the separation of timescales prohibits multiple transitions, $q(\vc{r}^N,t)$ reduces to the time independent committor function of transition path theory.\cite{das2021variational,helfmann2020extending}  The explicit relation between the Doob force $\boldsymbol{\lambda}^*(\mathbf{r}^N,t)$ and $q(\mathbf{r}^N,t)$ is,
\begin{equation}\label{eq:1.10}
    \boldsymbol{\lambda}^*(\mathbf{r}^N,t) = 2\kbt \nabla \ln q(\mathbf{r}^N,t)
\end{equation}
where by construction this force makes all trajectories reactive, and the reactions occur as they would in the original system. This force uniquely saturates the inequality in Eq. \ref{eq:1.9}, thus providing a unique description of the reaction in a complex system.\cite{das2022direct}

This formalism allows us to compute both the time-dependent committor $q(\mathbf{r}^N,t)$ and the rate $k$ from a reactive trajectory ensemble by parameterizing the external force $\boldsymbol{\lambda}$ and optimizing it by maximizing the expectation value of the change in action averaged within the reactive trajectory ensemble. In this work, we will consider parameterizing $\boldsymbol{\lambda}$ with both linear functional forms as well as a non-linear form provided by a neural network. The optimization of either is done by defining a loss function, $\mathcal{L}_\lambda$, as
\begin{align}\label{eq:1.12}
    \revision{\mathcal{L}_\lambda = \bigg{\langle}  \sum_{n=0}^{t_f/\Delta t} \sum_{i=1}^N \frac{\Delta t}{4\gamma_i k_{\mathrm{B}}T} \big{(} -|\boldsymbol{\lambda}_{i}(n \Delta t)|^2} \\ \notag   
    \revision{+ 2 \boldsymbol{\lambda}_i(n\Delta t)\cdot\boldsymbol{\eta}_i(n \Delta t) \big{)}  \bigg {\rangle}_{B|A}}
\end{align}
where the \revision{sum is over each particle that the noises act on}, and $\boldsymbol{\lambda}_i$ is the component of the driving force on the degree of freedom associated with the noise. This loss function is just the change in stochastic action \revision{in the discretized form}, so in optimizing $\boldsymbol{\lambda}_i$ we are simultaneously optimizing our estimate of the rate in the reference system. This optimization occurs over $n_\mathrm{int}$ iterations, and requires averages within the reactive trajectory ensemble, which we will generate with standard path sampling tools like transition path sampling. \revision{Since this method requires the positions and the noises at each time step, the method of generation of transition path ensemble requires some care. For complex systems where we use transition path sampling, we store the random number seeds for each runs, and then save the new trajectories only when they are accepted by rerunning them with the same seed. We find this method to be minimally slower than running standard transition path sampling. The option to save and change the seeding is available in most molecular dynamics simulation packages.}

\section{Choice of Ansatz and Convergence}
The accuracy of the rate estimate, and the mechanistic information afforded by the evaluation of $q(\mathbf{r}^N,t)$, depends on the fidelity with which the function can be represented. This depends on the ansatz used to expand it, and in particular, its expressibility. It also depends on the ease by which the function is learned, as inevitably the reactive path ensemble needed to train $q(\mathbf{r}^N,t)$ will be computationally expensive to generate. In this section, we consider the relative merits of expanding the driving force in both linear and nonlinear bases, and assess their accuracy and data efficiency. 

\subsection{Linear Function Ansatz }
\begin{figure}[t]
  \centering
    \includegraphics[trim={0.6cm .7cm .4cm 0.5cm},clip,width=8.5cm]{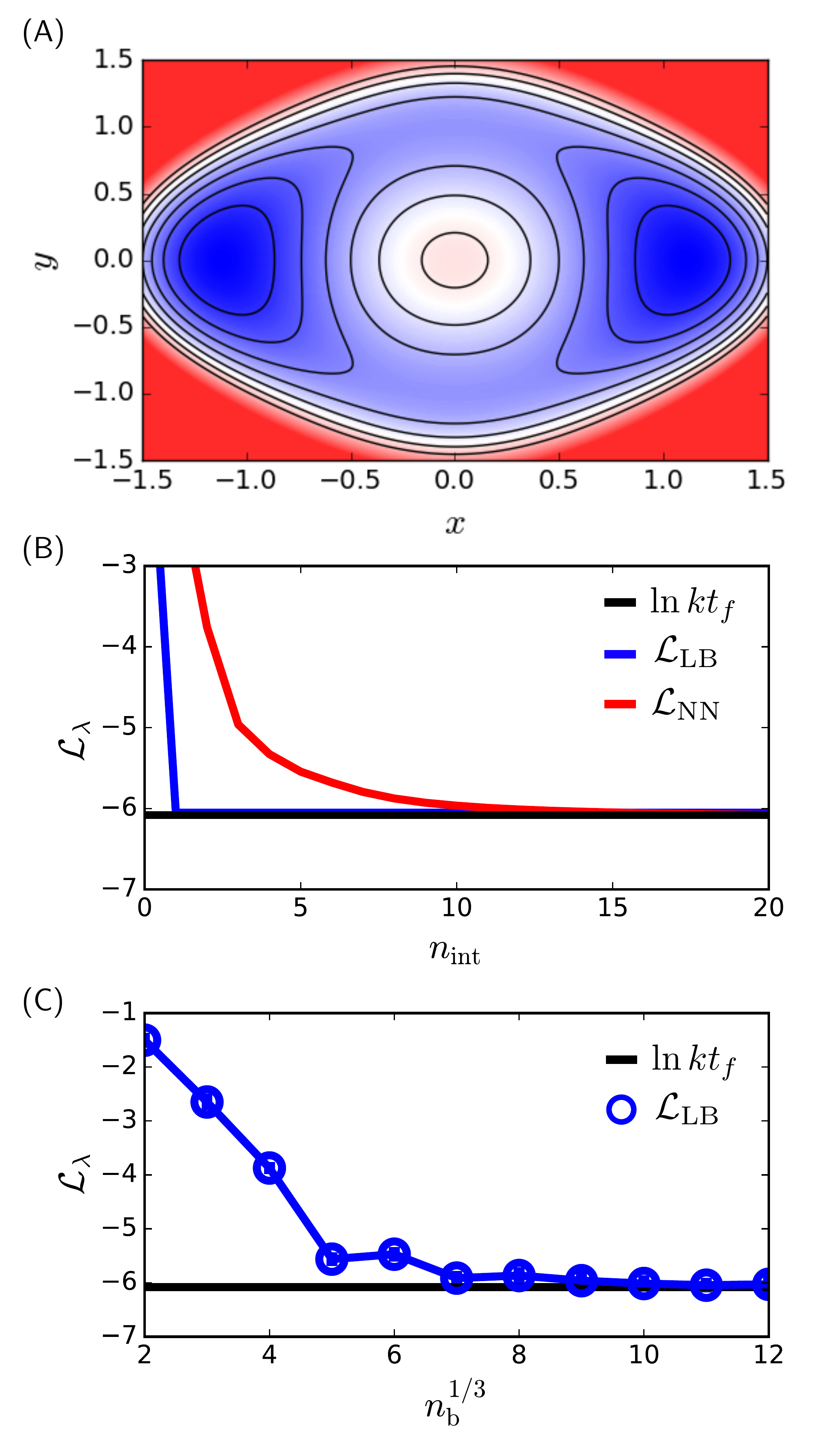}
    \caption{Functional ansatz testing. (A) Potential energy surface of the simple 2D model where the lines denote a separation of 2 $\kB T$. (B) Convergence of the loss function using the linear ($\mathcal{L}_\mathrm{LB}$) ansatz and neural-net ($\mathcal{L}_\mathrm{NN}$) ansatz. (C) Convergence of the linear basis with basis set size. Error bars denote one standard error computed from 3 independent trials.}
    \label{fig:1}
\end{figure}

We first consider the case of linear function approximations. A linear functional for $\boldsymbol{\lambda}(\vc{r}^N,t)=2\kB T \nabla \ln q(\mathbf{r}^N,t)$ can generically be expressed as its associated potential, 
\begin{equation}
2 \kbt \ln q(\mathbf{r}^N,t) = \sum\limits_{n=1}^{n_\mathrm{b}} c_n \varphi_n(\vc{r}^N,t) 
\end{equation}
where $c_n$ and $\varphi_n(\vc{r}^N,t)$ denote the $n$'th coefficient and basis function, and $n_\mathrm{b}$ denotes the total number of basis functions. This can be written compactly as $\boldsymbol{\lambda}(\vc{r}^N,t)=\nabla [\mathbf{c}\cdot \mathbf{\Phi}(\vc{r}^N,t)]$, where $\mathbf{c}$ is the $n_\mathrm{b}$ length vector of coefficients and $\mathbf{\Phi}(\vc{r}^N,t)$ is the vector of basis functions. For a linear functional expansion, the optimal set of coefficients $\mathbf{c}^*$ has a closed form that can be computed by taking the derivative of the loss function in Eq. \ref{eq:1.12} and setting it to 0. Computing the coefficients reduces to solving a $n_\mathrm{b} \times n_\mathrm{b}$ set of linear equations, whose solution is
\begin{equation}\label{eq:2.1}
 \mathbf{c^*} = \left[ \bigg{\langle}   \int_0^{t_f} dt\; \nabla \mathbf{\Phi} \otimes \nabla \mathbf{\Phi} \bigg{\rangle}_{B|A} \right]^{-1}  \bigg{\langle} \int_{0}^{t_f}dt\; \boldsymbol{\eta} \cdot \nabla \mathbf{\Phi}  \bigg{\rangle }_{B|A}
\end{equation}
where $\otimes$ denotes an outer product. For an orthonormal basis, the optimal coefficients are simply related to the average noise-weighted basis function\cite{dolezal2022mechanical}, but in general, the functions are not expected to be orthonormal. Because of this simplicity in training, linear bases are particularly efficient to employ. In cases where the reaction coordinate can be described well by a limited set of coordinates or order parameters, they can also be accurate\cite{das2019variational,das2021reinforcement,das2022direct,strahan2021long,thiede2019galerkin}. 

To understand the utility of a linear functional approximation, we consider a particle evolving in a two-dimensional external potential with two reactive channels visualized in Fig. \ref{fig:1} (A). The potential $V(x,y)$ is 
\begin{equation}
V(x,y)/\kB T = 2 [6 + 4 x^4 - 6 y^2 + 3 y^4 + 10 x^2 (y^2-1) ]
\end{equation}
where $x$ and $y$ are dimensionless coordinates and we have worked in a reduced unit system determined by $\kB T = \gamma_x = \gamma_y = 1$, and employed a first order Euler integrator with timestep equal to 0.004 $t^*$ with \revision{$t^*=\gamma_x/\kbt$} as our reduced time unit. We considered transitions defined by the indicator functions 
\begin{equation}
\revision{h_A(t) = \Theta(-x(t)+0.85) \qquad  h_B(t) = \Theta(x(t)-0.85)}
\end{equation}
where $\Theta$ denotes the Heaviside step function. A reactive path ensemble was generated by running brute force trajectories in order to sample 400 reactions, and the rate was evaluated by computing the side-side correlation function. We found that $t_f/t^*=2$ was a sufficient observation time to be in the linear growth regime for the transition probability with $\ln k t_f = -6.1\pm 0.1$. 

The linear approximation used were localized Gaussian basis functions of the form
\begin{equation}
\varphi_n(x,y,t) = e^{-a_x(x-x_n)^2}e^{-a_y (y-y_n)^2}e^{-a_t (t-t_n)^2}
\end{equation}
where  the Gaussian centers $\{x_n, y_n, t_n \}$ were equally spaced on a grid within the range of $x=[-1.5,1.3]$, $y=[-1.6,1.6]$ and $t=[0,2]$ and
Gaussian widths were choosen such that $\{a_x=1.4/(n_\mathrm{b}^{1/3}-1), a_y=1.6/(n_\mathrm{b}^{1/3}-1), a_t=1/(n_\mathrm{b}^{1/3}-1) \}$. The expansion coefficients, $\mathbf{c^*}$, were computed using Eq. \ref{eq:2.1} averaged over the path ensemble consisting of the 400 reactive trajectories. The optimization was done in one step, $n_\mathrm{int}=1$ \revision{by solving the linear equation in Eq. \ref{eq:2.1}}, where we found the loss function immediately converged to the brute force estimate of the rate for $n_\mathrm{b}=12^3$ as shown in Fig. \ref{fig:1} (B). The dependence of the rate estimate with the size of the basis is shown in Fig. \ref{fig:1} (C), where the loss decays slowly, obtaining a value of the rate statistically indistinguishable from the brute force estimation of the rate for \revision{$n_\mathrm{b}=8^3$}. This slow decay could be mitigated somewhat with fine tuning the basis, but we do not explore that here.

\subsection{Neural Network Function Ansatz}
Since the form of the force is rapidly varying and nonlinear, saturation of the inequality in Eq. \ref{eq:1.9} requires a large number of basis functions. If we express the linear ansatz in the full configuration space, the number of basis set coefficients grows exponentially with the degrees of freedom, making it intractable to converge the loss to the rate for complex systems. In order to circumvent the exponential scaling of the number of basis sets with the dimensionality of the system,  we consider employing a neural network (NN) ansatz to compute the time dependent committor, associated Doob force through automatic differentiation, and evaluate the rate through optimization. The input comprises the features selected for expressing the force and is connected to two hidden layers. For the two hidden layers, the Swish activation function\cite{ramachandran2017searching} is used as its derivatives are free from discontinuities, while also being exempt from the weight decay problem.\cite{montes2021differentiable} The penultimate layer only contains a single unit, with a sigmoid activation function. The output of this layer is the model's estimate of the time-dependent committor, $q(\vc{r}^N,t)$. The final layer is a \emph{lambda} layer, which simply computes the log of the committor. The output of this layer represents the many-body potential, $\ln q(\vc{r}^N,t)$, and the forces can be computed by taking a derivative of the output with respect to the input coordinates via autodifferentiation. While one can simply parametrize the forces instead of the committor, this architecture automatically enforces the conservativeness of the potential and offers a simple way to obtain the committor without the need to perform a multidimensional integration.

For the NN ansatz for the same 2D system above, $x$, $y$ and $t$ were used as the input features and optimization was performed using the RMSprop optimizer\cite{hinton2012neural} on 200 reactive trajectories. The learning rate was choosen to be 0.001, and for each iteration the loss function and associated gradients were evaluated over half of the trajectories drawn randomly from the ensemble. The training curve plotted in Fig. \ref{fig:1} (B) shows that the loss function plateaus to $\ln kt_f$ within $n_\mathrm{int}=20$ indicating that this ansatz was successful in learning the exact time-dependent committor quickly. While the training required multiple iterations, the number of parameters used to converge to the brute force rate was around 500 without specific optimization, fewer than required in the naive linear function approximation. The flexibility of the NN ansatz and the relatively swift training suggest it as a viable means of approximating the time dependent committor. As a consequence, in the remainder of the manuscript, we consider only the performance of the NN ansatz. 

\subsection{Convergence with Limited Statistics}

To illustrate the efficiency of this method, we tested the convergence of the NN ansatz with the statistics used to compute the rate and time dependent committor within the previously \revision{introduced} model two-dimensional potential. 
Specifically, we tested the convergence of the NN ansatz with the number of reactive trajectories used in training, as well as the time lag between configurations along a reactive trajectory. For both cases, we use two estimators, one which probed how closely the restricted trajectory ensemble is to the full trajectory ensemble, and a second which indicated how well a model trained on an approximated trajectory ensemble performs on the original trajectory ensemble. \revision{All of these estimators are based on comparisons between the OM action and $\ln kt_f$, as the agreement between the two signifies the success of the model in learning the true committor. Hence, the difference between the two offers a natural metric to probe the error between the parameterized and the true committor.} We denote the error from each of these approximate estimates as $\Delta \mathcal{L}$.

\begin{figure}[t]
  \centering
    \includegraphics[width=\linewidth]{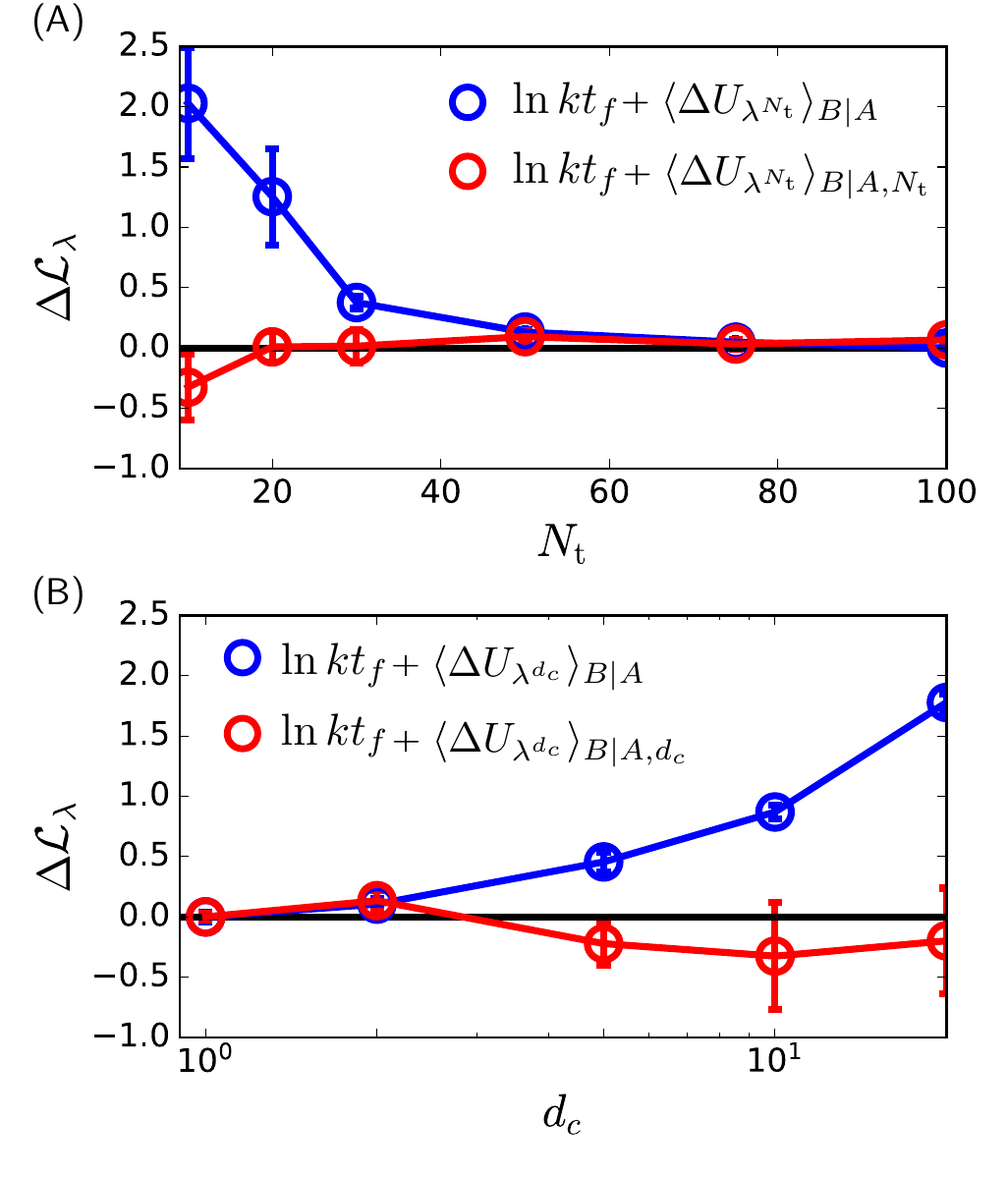}
    \caption{Convergence of rate estimates with respect to the accuracy of the reactive trajectory ensemble.  (A) Error estimators for the loss function of the NN ansatz as a function of the number of reactive trajectories $N_\mathrm{t}$ used for training. (B) Error estimators for the loss function of the NN ansatz as a function of the number of configurations used per trajectory. $d_{c}=1$ corresponds to the original ensemble, where every configuration is used for training. \revision{Error bars} denote one standard error computed from 5 independent trials.}
  \label{fig:2}
\end{figure}

For the first case, we vary the number of trajectories $N_\mathrm{t}$ used for training the model. The model trained on this limited trajectory ensemble, with force denoted by $\lambda^{N_\mathrm{t}}$ is then used to compute the first cumulant for the original trajectory ensemble comprised of the full 200 trajectories, $\langle \Delta U_{\lambda^{N_\mathrm{t}}} \rangle_{B|A}$. This is compared to the first cumulant obtained by training the model on the original trajectory ensemble with an optimal estimate of the rate. The difference of these two values, plotted in Fig. \ref{fig:2} (A), is an indicator of how close the committor trained on the restricted ensemble is to the actual committor. This plot shows that the estimator converges quickly with $N_\mathrm{t}$, and suggests that about 50 trajectories are sufficient to learn the time-dependent committor for this specific system. Another way of probing how the restricted trajectory ensemble compares to the original trajectory ensemble is to perform both training and averaging in the restricted trajectory ensemble, and compare that estimate to the true rate. This difference between the action averaged in a restricted ensemble $\langle \Delta U_{\lambda^{N_\mathrm{t}}} \rangle_{B|A ,N_\mathrm{t}}$, also shown in Fig \ref{fig:2} (A), is observed to be negative for $N_\mathrm{t}=10$, indicating overfitting of the model to the restricted trajectory ensemble. However, this error vanishes quickly, and plateaus to 0 for $N_\mathrm{t}=50$. This is a reflection of the transition path ensemble and the similarity of different reactive trajectories. The \revision{error bars} for all of these cases are obtained by training 5 different models on $N_t$ randomly selected trajectories from the  trajectory ensemble.

For the second case, we approximate the trajectory ensemble by storing only every configurations after a timelag of $d_c \Delta t$ where $\Delta t$ is the timestep used to integrate the trajectory. The original reactive trajectory ensemble comprises 200 trajectories with 500 discrete timesteps, and the number of configurations used per trajectory are obtained by dividing 500 by $d_c$. We train the model in this approximated trajectory ensemble and compute the same error estimates. The first estimate compares the first cumulant averaged in the original trajectory ensemble ($d_c=1$) with the model trained on the approximated trajectory ensemble to the loss computed by performing both averaging and training in the original trajectory ensemble, $\langle \Delta U_{\lambda_{d_c}} \rangle_{B|A}$. In this case, this estimate probes how well the the NN ansatz is able to extrapolate the forces for timesteps that it has not been trained on. The difference plotted in Fig. \ref{fig:2} (B) indicates that this extrapolation fails quickly. The second estimate probes the loss obtained by performing both averaging and training in the approximated trajectory ensemble whose loss is $\langle \Delta U_{\lambda_{d_c}} \rangle_{B|A, d_c}$. To get this estimate, the variance in Eq. \ref{eq:1.12} had to be scaled by a factor of $d_c^{-1}$ to account for the change in the effective timestep $\Delta t$. This difference plotted in Fig. \ref{fig:2} (B) shows that this approximation only works well for $d_c \leq 5$. The poor scaling with $d_c$ reflects the fact that stochastic diffusions with different variances have no overlap in the continuum limit.\cite{oksendal2013stochastic} 

\section{Rate decomposition and feature selection}
From an information theoretic point of view, the rate is a ratio of a conditioned and an unconditioned trajectory partition function\cite{bolhuis2002transition,dellago1999calculation,dellago2006transition,louwerse2022information}. Our optimization directly minimizes the KL-divergence between a trajectory ensemble driven with force $\boldsymbol{\lambda}$ and the undriven reactive trajectory ensemble. As the KL-divergence is expressible by the change in stochastic action along a trajectory, it involves a sum over all the degrees of freedom that the noises act on. For a suboptimal force, the rate is given by the average of the exponential of this quantity, coupling the noises from different degrees of freedom. However, when the variational bound is saturated and the rate is given by a simple mean, the accompanying change of action is linearly decomposable. This decomposition provides mechanistic insight, and affords a means of optimizing the features that form the representation of $\boldsymbol{\lambda}$. We generally find a NN ansatz to saturate the bound in Eq. \ref{eq:1.9}, which allows us in this section to explore a variety of featurizations and their corresponding contributions to the rate. Specifically, we consider networks with Cartesian and collective coordinates, as well as those integrated with underdamped equations of motion.

\subsection{Cartesian coordinates}
Using an NN ansatz allows us to compute the exact time dependent committor and associated Doob force. When Eq. \ref{eq:1.9} is saturated, the rate is given by the first cumulant of the change in action. This allows us to decompose the rate into independent contributions,
\begin{align}\label{eq:3.1}
    & kt_f = \exp\left [- \sum_{i=1}^{Nd} \langle \Delta U^i_{\lambda^*}\rangle _{B|A} \right ]
\end{align}
where
\begin{align}
 \Delta U^i_{\lambda^*} &=  \int_0 ^{t_f} dt \frac{[\lambda^*_i(t)]^2}{4 \gamma_i \kbt} 
\end{align}
 is the contribution to the rate per stochastic degree of freedom. The change in action, $ \Delta U^i_{\lambda^*} $, is strictly positive, indicative of the transition probability being less than 1, and results from functional minimization of Eq. ~\ref{eq:1.12}. The stochastic action for the Langevin equation is a sum of Gaussian random variables for each degree of freedom at each timeslice, and a change in stochastic action is a difference of Gaussian random variables. Given this, and recognizing the quadratic dependence on $\lambda^*_i(t)$ for the change in action, we observe that $\lambda^*_i(t)$ is essentially fitting the bias in the Gaussian noises generated when conditioning the stochastic process to react. Therefore, only degrees of freedom that require activation, or a rare sequence of noises, will accumulate a significant change in stochastic action or contribute significantly to the rate. Degrees of freedom that are uncorrelated with the reaction will not contribute to the rate, as their noises will remain unbiased. 

\begin{figure}[t]
  \centering
    \includegraphics[trim={0.13cm 0.35cm .25cm 0.10cm},clip,width=\linewidth]{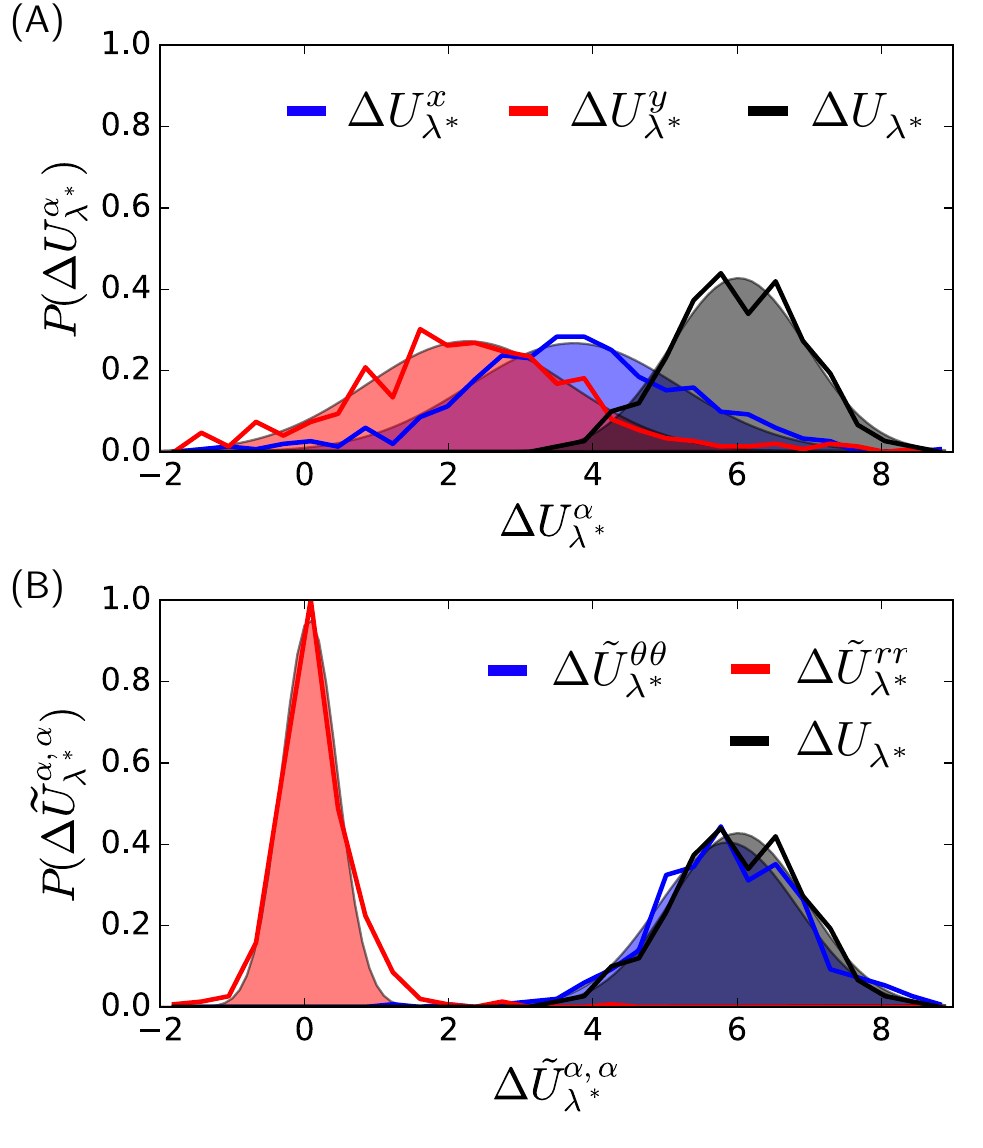}
    \caption{Decomposition of the rate into contributions from different degrees of freedom. (A) The distribution of the relative action for the two Cartesian coordinates $x$ and $y$.  (B) The decomposition of the rate performed for the polar coordinates $r$ and $\theta$, using Eq. \ref{eq:3.1}. Due to the orthogonality of the transformation, and the isotropicity of the diffusivities, the coupling term is zero. The perfect overlap between the $\Delta \tilde U_\lambda^{\theta \theta }$ and the total relative action identifies $\theta$ to be an excellent descriptor of the reaction coordinate.}
  \label{fig:3}
\end{figure}

To illustrate how this decomposition can be used to identify the relevance of coordinates, we consider the same 2D system visualized in Fig. \ref{fig:1} (A) and perform a decomposition of the rate. The two stochastic coordinates, $x$ and $y$, are fed into the neural network \revision{ansatz} and optimized. The resultant distributions for the individual components of the stochastic action, $P[\Delta U_{\lambda^*} ^\alpha]$, for $\alpha=\{x,y\}$, defined as
$$
P[\Delta U_{\lambda^*} ^\alpha] = \left \langle \delta( \Delta U_{\lambda^*} ^\alpha-\Delta U_{\lambda^*} ^\alpha[\vc{X} ])\right \rangle_{B|A}
$$
are shown in Fig. \ref{fig:3} (A). Neither of the two distributions show a complete overlap with the distribution of the total rate, $P[\Delta U_{\lambda^*}]$, indicating that both $x$ and $y$ are important in describing the reaction coordinate. However $P[\Delta U_{\lambda^*} ^x]$ is shifted towards larger values, and the expectation value of $\langle \Delta U_{\lambda^*}^x \rangle_{B|A}$ is found to be be larger than $\langle \Delta U_{\lambda^*}^y \rangle_{B|A}$, allowing us to quantitatively assert that the coordinate $x$ is more important to the reaction than $y$, as it encodes more information of the conditioned path ensemble. However $y$ is still relevant, in agreement with intuition from the geometry of the potential.

\subsection{Collective coordinates}
While the decomposition above can quantify the relevance of an coordinate to a reactive event, they are expressed in the bare Cartesian coordinates that enter into the equation of motion. As such, their utility is diminished in many-particle systems which are translationally and rotationally invariant, and for which the number of degrees of freedom is large. A canonical approach in the study of rare events in complex systems is to employ collective coordinates, which are nonlinear combinations of the original Cartesian coordinates and may encode the expected symmetries of the system. In order to extend the formalism into this regime, we consider the transformation between Cartesian and collective coordinates, $\vc{r} \rightarrow \tilde{\vc{r}}$, and its subsequent impact on the rate decomposition. The Jacobian of the transformation $\mathbf{J}_{\boldsymbol{r}}(\tilde{\boldsymbol{r}})$ is,
$$
\mathbf{J}_{\boldsymbol{r}}(\tilde{\boldsymbol{r}}) =
 \begin{bmatrix} \nabla_{\boldsymbol{r}} \tilde{\vc{r}}_1 & \cdots & \nabla_{\boldsymbol{r}} \tilde{\vc{r}}_{\tilde n} \end{bmatrix}
$$
which is a matrix of $Nd \times \tilde{N}$ partial derivatives where $\tilde{N}$ is the size of the collective variable function space. Under this transformation, the original forces $\boldsymbol{\lambda}(\mathbf{r}^N,t)$ and the transformed forces $\boldsymbol{\tilde \lambda}(\mathbf{\tilde{r}},t)$ are related by
\begin{equation}\label{eq:3.2}
    \boldsymbol{\lambda}(\boldsymbol{r}^N,t) = \mathbf{J}^{T}_{\boldsymbol{r}}(\tilde{\boldsymbol{r}})\cdot \boldsymbol{\tilde\lambda}(\boldsymbol{\tilde r},t) 
\end{equation}
where the force acting on the original coordinate $\vc{r}_i$ due to the force $\boldsymbol{\tilde{\lambda}}_j$ which depends on the collective coordinate $\tilde{\vc{r}}_j$ is given by a product of  $\tilde{\boldsymbol{\lambda}}_j$ and the Jacobian element $J_{ij}$. Inserting this into the expression for the optimal stochastic action, in Eq.~\ref{eq:3.1}, we obtain
\begin{align}
    \Delta U_{\lambda^*} &= \sum_{j,k}^{\tilde{N}} \Delta \tilde U^{jk}_{\lambda^*}
\end{align}
with
\begin{align}\label{eq:3.3}
   \Delta \tilde U^{jk}_{\lambda^*} =  \int_0^{t_f} dt \frac{\tilde \lambda_j^*  \tilde \lambda_k^*}{4k_\mathrm{B} T} \Gamma^{-1}_{jk} 
\end{align}
where the initially linearly independent factors from each Cartesian coordinate, indexed by $i$, are expressed as a pair of contributions from the collective coordinates, indexed by $j$ and $k$. From this form it is evident that the contribution to the rate incurred from the transformed coordinates $\vc{\tilde{r}}$ are not necessarily independent of each other \revision{or bipartite}\cite{louwerse2022information}. Zero coupling between $\tilde r_j$ and $\tilde r_k$ is obtained when the effective friction $\Gamma^{-1}_{jk}=\sum_{i=1} J_{ij} J_{ik} / \gamma_i = \delta_{jk}/\gamma$, with $\gamma_i=\gamma$, a condition that requires the friction weighted transformed coordinates to be orthogonal.

As an illustration of the decomposition under a change of coordinates, we consider the same 2D system as before, but rather than parameterizing $\boldsymbol{\lambda}_i$ on $x$ and $y$, we transform into polar coordinates $(x,y) \rightarrow (r,\theta)$, where $r = x^2 + y^2$ and $\tan \theta = y/x$. 
We quantify the contributions to the rate from the polar coordinates, by training the NN ansatz on the polar coordinates. The partials are prepared ahead of time and are passed into the loss function along with the noises. The relative action distributions in the transformed coordinates, $P[\Delta \tilde{U}_{\lambda^*} ^{\alpha,\alpha'}]$ for $\alpha,\alpha' = \{r,\theta\}$ computed using Eq. \ref{eq:3.3} are shown in Fig. \ref{fig:3} (B). Since polar coordinates are orthogonal and $\gamma_x=\gamma_y$, the coupling term $\Delta \tilde{U}_{\lambda^*} ^{\alpha,\alpha'} =0$ for $\alpha \ne \alpha'$. We observe that the distribution corresponding to the coordinate $\theta$ almost perfectly overlaps with the total action distribution, indicating that $\theta$ is an excellent descriptor of the reaction coordinate. The distribution for $r$ is centered around 0 and narrow, illustrating it is unbiased by conditioning on a reaction and thus contributes little to the rate. This decomposition of the rate in collective coordinates provides a simple metric to identify the relevance of physically meaningful descriptors to a reactive process, without making any \textit{a-priori} assumptions about the reaction. The form of this metric is purely based on the physical mechanism of the reaction, as it quantifies how conditioning a trajectory ensemble to be reactive shifts the noise distributions per-coordinate. This allows us to do hypothesis testing for the relevance of collective coordinates, and discover the coordinates that are gating the rare event, and those that are uncorrelated with barrier crossing, through the size of their contribution to the rate. This hypothesis testing requires the saturation of the variational bound, which if not achieved points to the lack of relevant features in the NN ansatz.

\subsection{Importance of velocity}\label{Sec:V}
In a general molecular system, motion is not overdamped and as a consequence the full phase space spanned by both configurational coordinates as well as their conjugate velocities are required to specify a reactive trajectory. In order to understand the importance of including velocity degrees of freedom in a parameterization of $\boldsymbol{\lambda}$, we consider formally when it can be neglected. For concreteness, we consider an the underdamped Langevin equation of the form
\begin{equation}\label{eq:4.1}
m \dot{\vc{v}}_i = -\gamma \vc{v}_i + \vc{F}_i(\vc{r}^N) + \boldsymbol{\eta}_i \qquad \dot{\vc{r}}_i = \vc{v}_i
\end{equation}
where $\vc{v}_i$ is the velocity of particle $i$ and the rest of the quantities are defined in the same way as in Eq. \ref{eq:1.1}. For simplicity we take the mass, $m$, and friction $\gamma$ to be independent of particle index, though generalizations are straightforward. 
We start by noting that the backward Kolmogorov equation takes the form,
\begin{equation}\label{eq:4.15}
    \partial_t q = -\sum_{i=1} \vc{v}_i \cdot \nabla_{\vc{r}_i}   q - \frac{\gamma \vc{v}_i}{m} \cdot \nabla_{\vc{v}_i} q + \frac{\vc{F}_i}{m} \cdot \nabla_{\vc{v}_i} q + \frac{2\gamma \kbt}{m^2} \nabla_{\vc{v}_i} ^ 2 q
\end{equation}
which when solved with the same boundary conditions as Eq.~\ref{Eq:OBKE} yields the time dependent committor function $q(\vc{r}^N,\vc{v}^N,t)$ whose arguments we suppress above for ease of notation. Since the noise acts only on the velocities, the Doob force is given by the gradient of the committor with respect to the velocities rather than the positions,
\begin{equation}\label{eq:4.2}
    \boldsymbol{\lambda}_i^*(\vc{r}^N,\vc{v}^N,t) = \frac{2\gamma\kbt}{m} \nabla_{\vc{v}_i} \ln q(\vc{r}^N,\vc{v}^N,t)
\end{equation}
thus naively it would seem that parameterizing a velocity dependence is crucial whenever an underdamped equation is used. However, in the limit that $\gamma^{-1} \rightarrow 0$, we find that the velocity dependence can be safely ignored. 

This can be understood via application of perturbation theory, where $q(\vc{r}^N,\vc{v}^N,t)$ is expanded in orders of $\gamma^{-1}$\cite{papanicolaou1976some,vanden2006transition,pavliotis2014stochastic}. To first order, $q(\vc{r}^N,\vc{v}^N,t)$ becomes
\begin{equation}\label{eq:4.3}
    q(\vc{r}^N,\vc{v}^N,t)= q_0(\vc{r}^N,t) + \frac{m \vc{v}}{\gamma} \cdot \nabla_{\vc{r}} q_0(\vc{r}^N,t) + \mathcal{O}(\gamma^{-2})
\end{equation}
where $q_0$ is independent of the velocity. Substituting the approximated form of $q$ into the underdamped backward Kolmogorov equation, we find,
\begin{align}\label{eq:4.5}
    \partial_t q &\approx  - \sum_{i=1}  \frac{\vc{F}_i}{\gamma}  \cdot \nabla_{\vc{r}_i} q_0- \frac{m \vc{v}_i^2}{\gamma} \nabla_{\vc{r}_i}^2 q_0+ \mathcal{O}(\gamma^{-2})
\end{align}
which when averaged over the Maxwell-Boltzmann distribution, yields
\begin{equation}
    \partial_t q = - \sum_{i=1}^N \frac{\vc{F}_i\left (\mathbf{r}^N \right )}{\gamma}\cdot \nabla_{\vc{r}_i} q_0-  \frac{\kbt}{\gamma}  \nabla^2_{\vc{r}_i} q_0 + \mathcal{O}(\gamma^{-2})
\end{equation}
which to first order in $1/\gamma$ is identical to Eq.~\ref{Eq:OBKE}, the overdamped backward Kolmogorov equation, with $q(\vc{r}^N,\vc{v}^N,t)\approx q_0(\vc{r}^N,t)$. As a consequence, the committor in the overdamped limit becomes a function solely of $\vc{r}^N$ and the Doob force is given by a gradient with respect to position. In Appendix A, we show that to $\mathcal{O}(\gamma^{-2})$ this approximation also saturates the variational inequality for the rate expression. 

\begin{figure}[t]
  \centering
    \includegraphics[width=\linewidth]{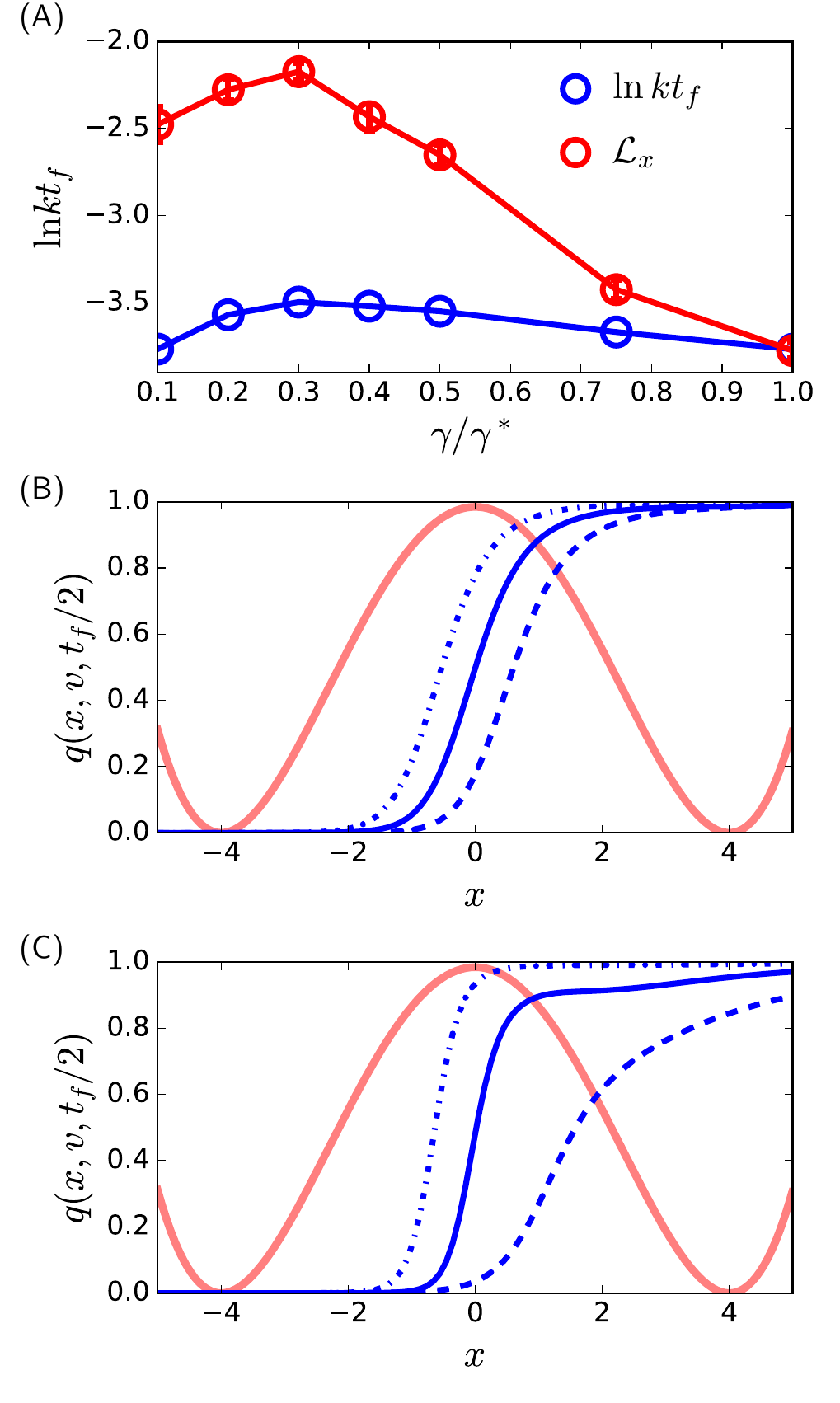}
    \caption{Computation of the committor for reactive processes that are integrated using the underdamped equations of motion. (A) The reaction rate and the error estimate in the loss function for optimizing the velocity-independent committor, $\mathcal{L}_x$, as a function of the friction coefficient $\gamma$. (B) and (C) show the optimized position and velocity dependent committor of the reaction between the metastable well depicted by the potential energy surface $V(x)$ in red, for friction coefficient $\gamma = 1.0$ and $\gamma  = 0.1$ respectively. For (B) and (C) the dot-dashed, solid and dashed lines denote slices of the committors at constant velocity, $v=-1,0$ and $1$, respectively.} 
  \label{fig:4}
\end{figure}

In order to gain intuition for when the higher order terms in Eq. \ref{eq:4.3} become negligible, we consider the reaction of a particle in a simple double well potential of the form
\begin{equation}
V(x)/\kB T = \frac{1}{64}(x-4)^2(x+4)^2
\end{equation}
where $x$ is a dimensionless coordinate and we take $\kB T = m = 1$, which determines a dimensionless time unit \revision{$t^*=\sqrt{m/\kbt}$}. We considered transitions between states defined by the indicator functions
\begin{equation}
\revision{h_A(t) = \Theta(-x(t)+3.6) \qquad h_B(t) = \Theta(x(t)-3.6)}
\end{equation}
and obtain 400 reactive trajectories each of length $t_f/t^*=5$ using a timestep of 0.01 $t^*$ using first order integrator. We studied this system over a range of $\gamma/\gamma^*$ between 0.1 and 1 with $\gamma^*=m/t^*$. We trained a NN ansatz only on the positions and time, and compared the optimized value of the loss function to the brute-force rates evaluated from a direct mean first passage time calculation. Figure \ref{fig:4} (A) shows the difference between the two estimates, along with the brute-force rate as a function of $\gamma/\gamma^*$. The reactive rates show a Kramers' turnover\cite{hanggi1990reaction} at $\gamma/\gamma^* \approx 0.3$, and the optimized loss is consistently off by a factor of 1.5 for $\gamma/\gamma^*<0.3$. After the turnover, the error in rate estimate decreases monotonically, until it completely vanishes for $\gamma/\gamma^*=1$. It is surprising that this relatively small friction is already consistent with the overdamped limit. 

To further understand the importance of velocity in the time dependent committor, we train a model to optimize for the velocity-dependent commitor for $\gamma/\gamma^*=0.1$ and $\gamma/\gamma^*=1$. For both of these cases, the optimized value of the loss was within a standard error of the true rate. The plot of the optimized committor evaluated at time $t=t_f/2$ as a function of velocity and position is shown in Figs. \ref{fig:4} (B) and (C). The functional dependence on time is not strong away from $t=0$ and $t=t_f$. For $\gamma/\gamma^*=1.0$, $q(x,v,t)$ depends weakly on $v$, with the dependence  being captured by a linear shift along $x$ to an otherwise simple sigmodal dependence on $x$. This is precisely the dependence expected from the expansion in Eq.~\ref{eq:4.3}. However, for $\gamma/\gamma^*=0.1$ the committor is strongly sensitive to the velocity. For $v=0$ the large $x$ behavior of $q(x,v,t)$ slowly converges to 1 reflecting the potential for the particle even at large values of $x$ to fail to react. For negative velocities, the inflection point of $q(x,v,t)$ is shifted to positive values of $x$, consistent with corresponding values of the potential that are low enough below the barrier that the particle is trapped. Correspondingly, for positive velocities, $q(x,v,t)$ is shifted to negative values of $x$, reflecting the high likelihood of reacting even for positions not quite to the top of the barrier. This behavior is not reproducible by scaling a spatially dependent committor by a simple constant. Hence, featurization of the velocity, or expansion of the committor to higher orders in $\gamma^{-1}$ is required to accurately encode the time-dependent committor for this low-friction regime.

\section{Application to Alanine Dipeptide}
To examine the efficacy of this method for a complex molecular system, we investigate the isomerization of alanine dipeptide.  Alanine dipeptide has two metastable conformations. It can transition between these two states via the rotation of the Ramachandran angles $\phi$ and $\psi$. A multitude of path sampling methods have focused on this model due to the collective nature of this transition in the gas phase and in solution. While the transition can be tracked using the $\phi$ and $\psi$, they serve only as order parameters and are not sufficient in describing the complete reaction coordinate or committor. \cite{bolhuis2000reaction,ma2005automatic} Significant advancements in methods to parameterize the time independent committor have been made by resolving this model along physically motivated, predetermined order parameters\cite{evans2022computing,mori2020learning,kikutsuji2022explaining,lopes2019analysis,li2019computing,elber2017calculating,ray2021markovian,gao2023transition,jung2019artificial,brotzakis2021method}. As we show, choosing among a large number of internal coordinates without consideration of their correlation or coupling risks neglecting important aspects of the transition path ensemble. This is because internal coordinates do not form an orthogonal set of coordinates, and collective motions such as the rotations of a single dihedral angle can be coupled with the motions of angles and other dihedrals. Below we first consider isomerization of alanine dipeptide in implicit solvent, and then in explicit solution. For both we parameterize $\ln q(\vc{r}^N,t)$ using the NN ansatz.

\begin{figure}[b]
  \centering
    \includegraphics[trim={0.3cm 0.1cm 0.35cm 0.1cm},clip,width=\linewidth]{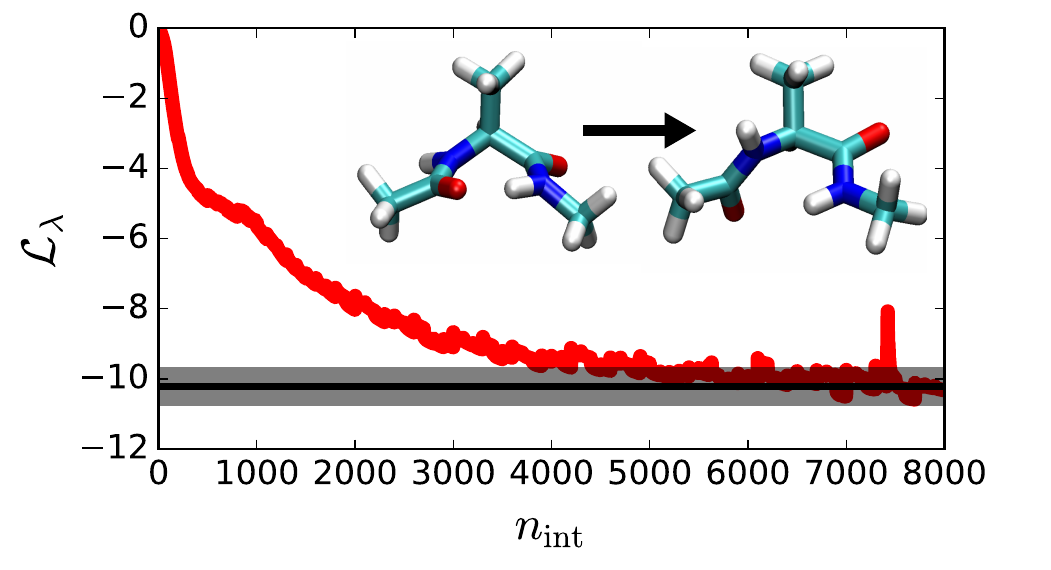}
    \caption{ Convergence of the loss function for isomerization in implicit solvent along with a representative snapshot of the two metastable conformations $\revision{C7}_\mathrm{ax}$ (left) and $\revision{C7}_\mathrm{eq}$ (right). Solid black line denotes the true rate, $\ln k t_f$, with shading denoting one standard error.} 
    \label{fig:5a}
\end{figure}

\subsection{Isomerization in implicit solvent}
In implicit solvent we consider isomerization of alanine dipeptide between its $\revision{C7}_\mathrm{eq}$ and $\revision{C7}_\mathrm{ax}$ conformations, as visualized in the inset in Fig. \ref{fig:5a}. 
To investigate this reaction, we first generated a reactive trajectory ensemble. Simulations were performed in OpenMM\cite{eastman2017openmm} and the AMBER ff14SB forcefield\cite{maier2015ff14sb} was used for parametrizing the dipeptide interactions. A Langevin thermostat with the leap-frog discretization was used as the integrator\cite{sweet2008normal}. The timestep was chosen to be 1 fs, $\gamma$ was set to $10$ ps$^{-1}$ and the transition path length $t_f$ was set as 1 ps. The indicator functions identifying the metastable wells $\revision{C7}_\mathrm{ax}$ and $\revision{C7}_\mathrm{eq}$ were defined using the  Ramachandran angle $\phi$,
\begin{equation}
\revision{h_A(t) = \Theta (\phi(t) - \pi/4) \qquad h_{B}(t) = \Theta (-\phi(t) + \pi/4)}
\end{equation}
and the first trajectory was generated by running forward and backward simulations from the top of the saddle point along the dihedral $\phi=0$. Transition path sampling\cite{bolhuis2002transition} was used to obtain a reactive trajectory ensemble and the shooting from the top method\cite{jung2017transition} was used to generate new trajectories. This method offers a way to decrease correlations between the trajectories as well as increase the acceptance rate for new trajectories by performing shooting moves within a restricted region near the saddle point, which for this case was chosen as $-\pi/6 \leq \phi \leq \pi/6$. A total of 1000 trial trajectories were generated, and the acceptance rate came out to be approximately 0.4. Every 5th trajectory in the ensemble was saved and used for analysis for a total of 200 trajectories. For the choice of reaction descriptors, we used all the internal coordinates that did not involve the hydrogen atoms. This set consists of 9 bonds, 11 angles and 12 dihedrals, the latter two of which contain 9 total redundant coordinates. These internal coordinates along with the Jacobians matrices are computed for the 200 saved trajectories, and saved to be used for training.

We use the underdamped approximation discussed in section Eq. \ref{eq:4.3}, which exempts us from including the velocities as a part of the feature set. The loss function is modified accordingly for the action of the Langevin leap-frog integrator\cite{izaguirre2010multiscale} implemented in OpenMM, as shown in Appendix B. The RMSProp optimizer with a learning rate of 0.001 was used to train the model, and training was performed for \revision{8000 steps. A 90-10 training validation split was used for optimization, and the splits were randomized every 100 epochs. Figure \ref{fig:5a} shows the value of the training set loss, along with the reaction rate obtained by computing the mean first passage time of 400 reactive trajectories generated independently. The loss function plateaus around the 4000th step, with the value being within one standard error of the true rate. }

\begin{figure}[b]
  \centering
    \includegraphics[trim={0.2cm 0.7cm 0.2cm 0.15cm},clip,width=\linewidth]{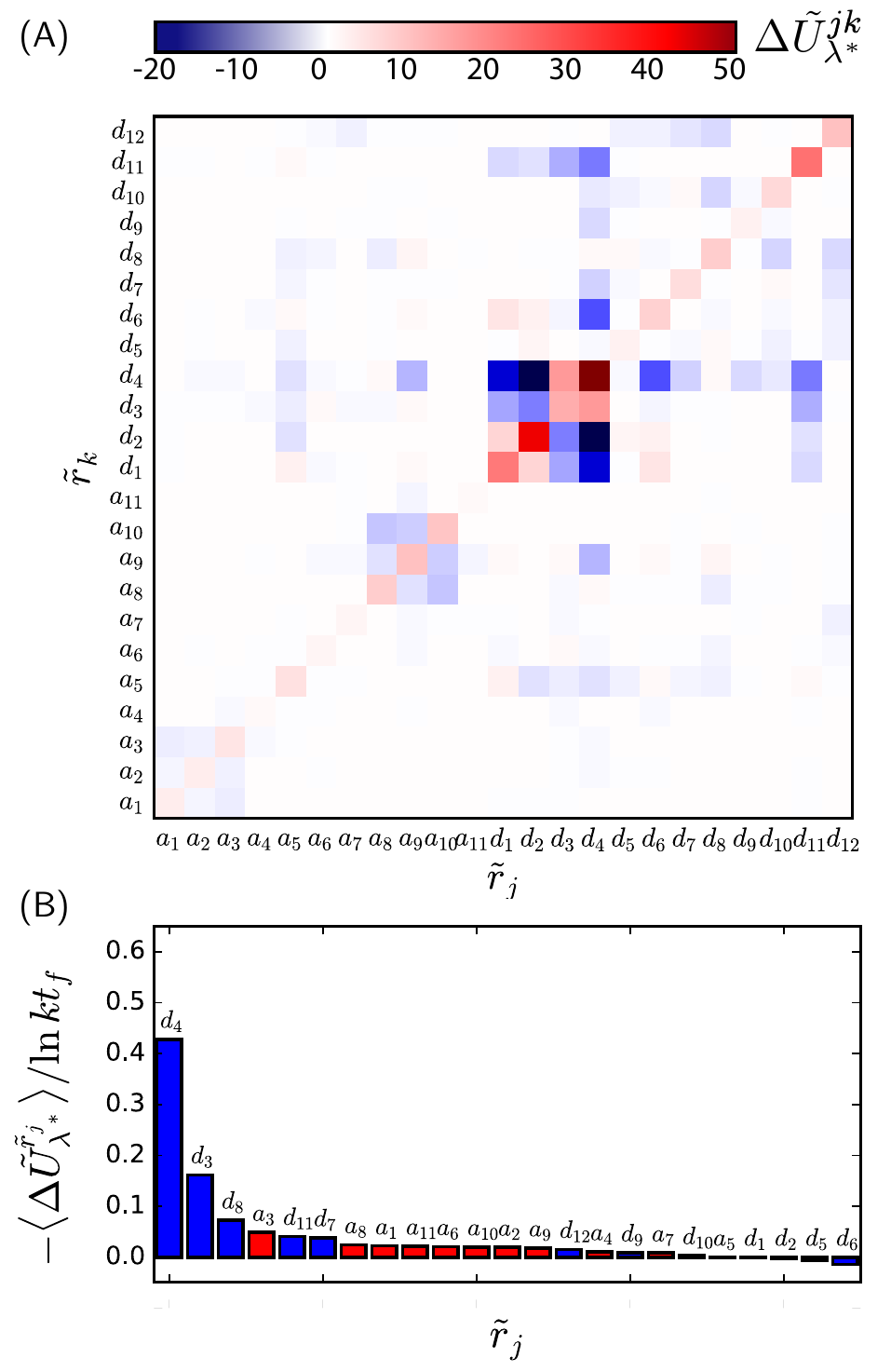}
    \caption{Decomposition of the rate of isomerization of alanine dipeptide in implicit solvent. (A)  The $\Delta \tilde{U}_{\lambda} ^ {jk}$ computed using Eq. \ref{eq:3.2} as a function of internal coordinates. (B) Decomposition of the rate in terms of contributions from internal degrees of freedom, computed by summing up the rows of the matrix in (A).} 
    \label{fig:5b}
\end{figure}

To gain mechanistic insight into the reaction, we performed the decomposition of the relative action in terms of the internal coordinates. Following Eq. \ref{eq:3.3}, we computed the $\langle \tilde U_{\lambda^*}^{jk} \rangle_{B|A}$ matrix and visualize it in Fig. \ref{fig:5b} (A). The angles and dihedrals are represented using the letters $a$ and $d$ respectively, and the numbers in the subscripts are defined in Appendix C. We observe that the matrix of contributions to the rate is sparse, with only a few select coordinates and their couplings obtaining a significant value. The contributions from the distances have been removed from the plot as their combined value was calculated to be statistically indistinguishable from zero. The effective decoupling of the bond vibrations from the angles and dihedrals provides evidence that the NN-based ansatz is not overfitting redundant features from the limited input dataset, consistent with physical intuition for stiff bonds.

We also observe a strong coupling among the dihedrals and the angles. Internal coordinates do not form an orthogonal set of coordinates, and the off-diagonal terms in the matrix indicate that the reaction is mediated by the coupling between these internal degrees of freedom. Moreover, the change in action is delocalized between sets of internal coordinates that have been ignored in previous studies. We note that the off-diagonal elements of the matrix are negative, while the diagonal terms are positive. This prevents us breaking down the rate in terms of additive contributions from different degrees of freedom. However, the matrix is symmetric so we can sum over the rows of the matrix and define the contribution from a single collective coordinate $j$ as
\begin{align}\label{eq:5.1}
    \Delta \bar U^{j}_{\lambda^*} = \sum_k^{ \tilde N} \Delta \tilde{U}^{jk}_{\lambda^*} 
\end{align}
where $\Delta \tilde{U}^{jk}_{\lambda^*}$ is defined the same way as in Eq. \ref{eq:3.3}. \revision{The reason for using this measure is based on the sum rule for the OM action along transformed coordinates, which from using calculus of variation follows}
\begin{equation}
    \revision{
    \tilde \lambda^*_j \sum_k^{\tilde N} \tilde \lambda^*_k \Gamma^{-1}_{jk} = \tilde \lambda^*_j \sum_{i}^{Nd} J_{ij} \eta_i}
\end{equation}
\revision{which is distinct from the case of Cartesian coordinates, where the relation is given by $[\lambda^*_i]^2 = \lambda_i^* \eta_i$. This difference reflects the fact that the noises along pairs of transformed coordinates are not necessarily independent of each other and cannot be assumed to follow bipartite dynamics\cite{louwerse2022information}. Summing over the rows of the matrix can be understood as a marginalization of the OM action over all the coupled coordinates}
Plotted in Fig. \ref{fig:5b} (B), this decomposition is found to be positive for almost all the internal coordinates except for two. These negative values are within the standard error. This allows us to extract the leading contributors to the $\revision{C7}_{\mathrm{ax}} \rightarrow \revision{C7}_{\mathrm{eq}}$ reaction. We observe that the Ramachandran angle $\phi$ ($d_4$) is found to incur the largest contribution. This is a remarkable result as no $\textit{a-priori}$ information of the reaction coordinate was passed into the model for training. While the indicator functions that were used to define the boundaries of the metastable wells were defined using $\phi$, the optimization scheme itself did not require any description of the indicator functions. Yet, this method automatically finds the Ramachandran angle $\phi$ to contribute the most to isomerization, out of 32 internal coordinates, 9 of which are redundant.

This decomposition reveals other leading contributors to the reaction and highlights other order parameters that are activated. The C-N-C$_{\alpha}$-C$_{\beta}$ ($d_3$) and the C$_{\beta}$-C$_{\alpha}$-C-N ($d_8$) torsions are found to be the next two leading contributors, suggesting that rotation of the Ramchandran angle $\phi$ is strongly coupled to the orientation of the alkyl bond. Some other important internal coordinates that are selected by this method include the O-C-N angle ($a_3$), the C-O-C-N improper torsion ($d_{11}$) and the C$_{\beta}$-C$_{\alpha}$-C-O ($d_7$) torsion. These internal coordinates also emphasize the importance of the relative orientation of the O-C bond and the methyl-bond. Our final observation is that the contribution from  other Ramachandran angle $\psi$ ($d_6$) is found to be effectively zero. This is another significant result as $\psi$ has long been used as the 2nd order parameter to explore the isomerization of alanine dipeptide due to the topology of the free energy surface. \revision{As this has been mentioned previously,\cite{bolhuis2000reaction} future studies should consider coarse-graining along other dihedral angles for performing committor analysis of this reaction.}

\subsection{Isomerization in Explicit Solvent}
\begin{figure}[t]
  \centering
    \includegraphics[trim={0.15cm 0.3cm .25cm 0.05cm},clip,width=\linewidth]{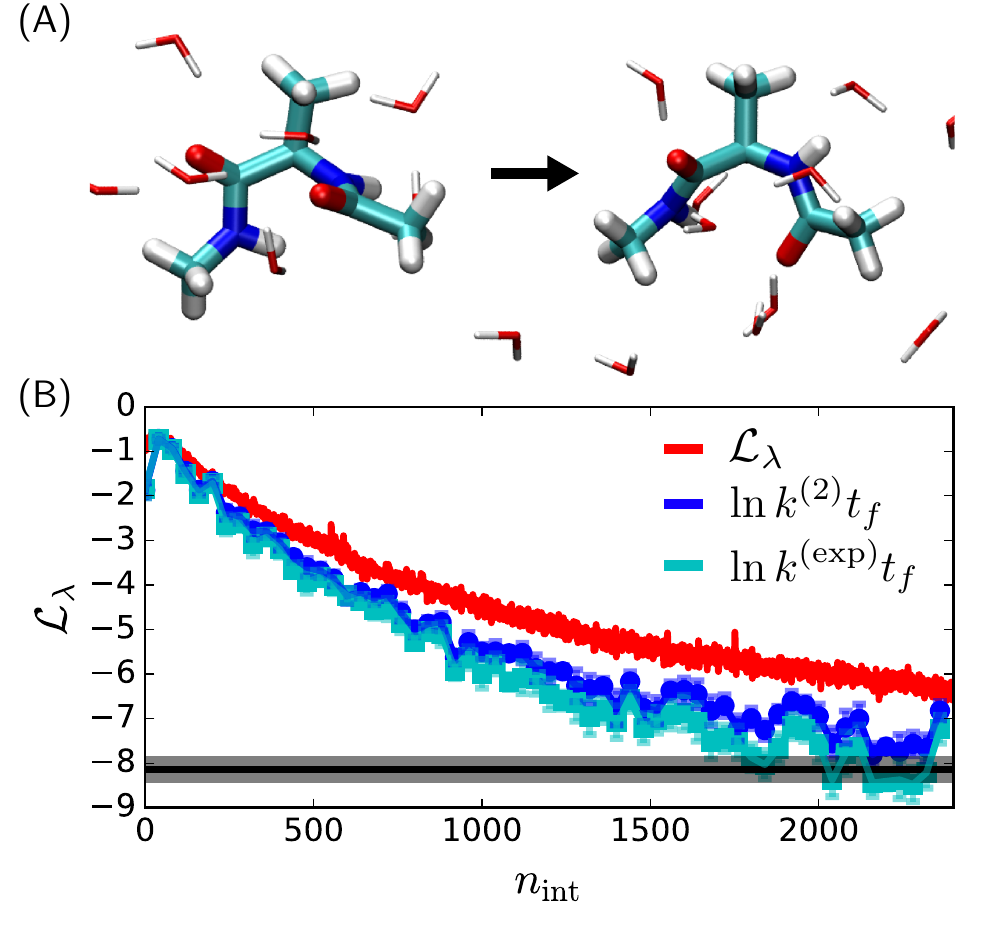}
    \caption{Investigation of the conformational isomerization of alanine dipeptide in explicit solvent. (A) Representative snapshot of the two metastable conformations, $\alpha_L$ (left) and $\beta$ (right). (B) Value of the loss function along with the two rate estimators $k^{(2)}$ and $k^{(\exp)}$ during training. The rate can be leveraged by the two estimators, even though the loss function does not converge to $\ln \; kt_f$. Solid black line denotes the true rate, $\ln k t_f$, with shading denoting one standard error.}
    \label{fig:6a}
\end{figure}

Finally to demonstrate the ability to tackle very high dimensional systems, we explore the conformational isomerization of alanine dipeptide in explicit solvent. As the potential energy landscape of along the Ramachandran angles is modified due to solvent interactions\cite{bolhuis2000reaction,vy}, we consider the isomerization between the $\revision{C5}$ and $\alpha_L$ states, visualized in Fig. \ref{fig:6a}. The equations of motion and forcefields for the peptide are the same as in the implicit solvent study, and the TIP3P forcefield\cite{jorgensen1983comparison} is used for parameterizing the water molecules. Lorentz-Berthelot mixing rules are used for the peptide-water interactions. A periodic box of volume 27 nm$^3$ is used with 862 water molecules, and the Ewald Summation is used for computing the long ranged interactions. The basin definitions for $\alpha_L$ and $\revision{C5}$ are the same as that of the $\revision{C7}_\mathrm{ax}$ and $\revision{C7}_\mathrm{eq}$ states, respectively. The same method as before is used for obtaining a reactive trajectory ensemble, with an ensemble of 200 reactive trajectories used for learning the time dependent committor.

For this reaction, we restrict the input feature to only parameterize the internal coordinates of the peptide. Noting that bond vibrations are decoupled from rotations of the dihedrals and that water interactions are mediated through hydrogen bonds, our input feature set comprises all the 36 angles and 45 dihedrals and contains 42 redundancies. While the solvent degrees of freedom can be parameterized using symmetry functions\cite{bartok2013representing,behler2007generalized,behler2011atom}, our goal is to illustrate how our method can leverage a quantitative insight into the reaction mechanism even when it does not have access to the full phase space. 

Optimization of the NN ansatz is performed using the RMSProp Optimizer for 2500 steps, and the results are shown in Fig. \ref{fig:6a} (B). The loss function plateaus to a value of 2 higher than $\ln k t_f$, \revision{which was computed through evaluation of the first mean passage time from 400 independently run trajectories. This discrepancy between the first cumulant and the true rate} is expected, as the feature set excludes the relevant solvent degrees of freedom. Regardless, we are able to obtain the correct rate from this method by computing the second cumulant and exponential average, the form of which is given in Eqs. \ref{eq:2.2} and \ref{eq:2.3}. Both these estimators are plotted, and are observed to converge to the rate computed independently. The agreement between the exponential estimator and second cumulant is only expected when the loss function is perturbatively close to the true value, otherwise the additional cumulants would be needed.  Note that even in the case where the Doob force is not fully optimal, driven trajectories are almost surely reactive. 

\begin{figure}[b]
  \centering
    \includegraphics[trim={0.35cm 0.4cm .30cm 0.05cm},clip,width=\linewidth]{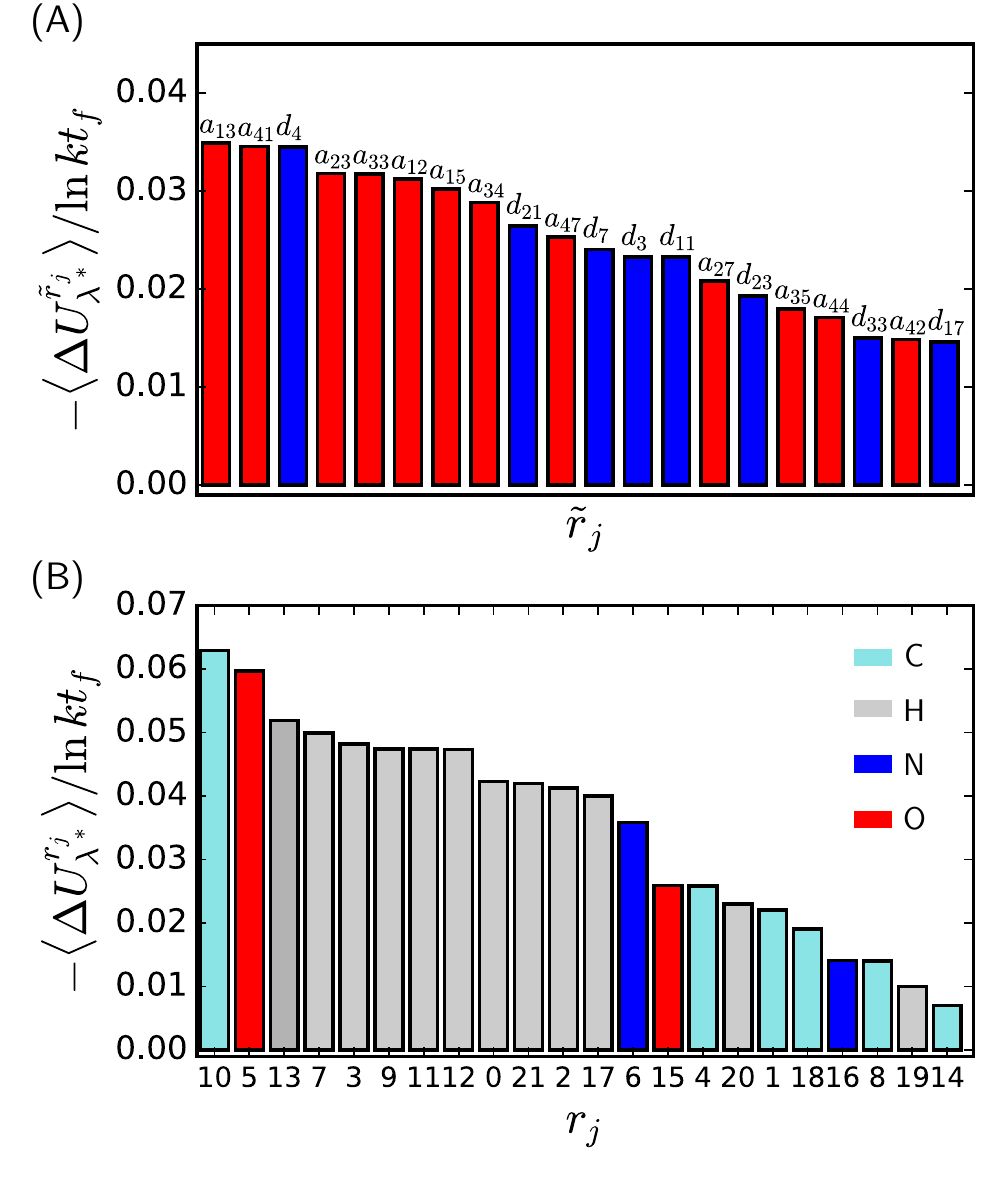}
    \caption{Decomposition of the rate of isomerization of alanine dipeptide in explicit solvent. (A) Decomposition of the rate in terms of contributions from the top twenty internal degrees of freedom, computed using Eq. \ref{eq:5.1}. (B) Decomposition of the rate in terms of contributions from all atoms, computed using Eq. \ref{eq:3.1}. The atom indices are labelled in the snapshot of the peptide in Appendix C. }
    \label{fig:6b}
\end{figure}
Since the variational bound is not saturated, the sum given in Eq. \ref{eq:3.1} does not equal the rate. This means that the description of the time dependent committor is not exact. However, because we are perturbatively close we can still use the method to extract the relative importance of degrees of freedom as before. First, we perform the same decomposition as Eq. \ref{eq:5.1}. While not visualized, the $\Delta \tilde{U}^{jk}_{\lambda}$ matrix is found to be less sparse for the reaction in solvent, due to the renormalization of the solvent effects into the peptide degrees of freedom. To gain quantitative insight, we sum of the rows of the $\Delta \tilde{U}^{jk}_{\lambda}$ matrix as before, and plot the contributions from the 20 leading features. Plotted in Fig. \ref{fig:6b} (A), this decomposition shows that the rotations of angles involving hydrogen atoms become more important than the internal rotations of the peptide dihedrals. Most of the leading contributors are angles that involve one or two hydrogen atoms, reemphasizing the effect of solvent in mediating this reaction. This is in accord with the findings of previous papers on the isomerization of solvated alanine dipeptide. \cite{ma2005automatic,velez2009kinetics,jung2019artificial,singh2022peptide}  However what is striking is that no single mode is dominant, with no internal coordinate accounting for more than  5\% of the rate.

To confirm the role of the hydrogen atoms, we plot a decomposition in terms of the individual atoms in Fig. \ref{fig:6b} (B) using the action expressed in the bare coordinates. The plot reveals that the methyl carbon contributes the most to the rate, followed by the acetyl carbonyl oxygen atom. However, the combined importance of the hydrogens far outweighs both.  This finding also illuminates why the addition of solvent transforms the reactive mechanism. Both these atoms strongly interact with water molecules via hydrophilic and hydrophobic effects that are mediated through hydrogen-bonding and volume exclusion, respectively.\cite{ma2005automatic,velez2009kinetics,jung2019artificial,singh2022peptide} We find that this method is able to provide a rate estimate and quantify the renormalized contributions from different degrees of freedom even when it does not have access to the full phase space. This feature can be particularly useful for more complex systems, where a complete description of the system is not tractable due to computational or memory bottlenecks.

\section*{Conclusion}
We have detailed a novel method that can be used to evaluate the time-dependent committor and the rate from a reactive trajectory ensemble. The method employs an ansatz for parameterizing a many-body potential that is related to the time-dependent committor, and can be optimized by variationally solving the backward Kolmogorov equation, as expressed through a trajectory reweighting theory used within variational path sampling. For reactive processes in equilibrium, where the cost of obtaining a reactive trajectory ensemble is independent of the rarity of the reaction, this method provides a simple procedure to compute the rate and distill mechanistic information.

Combining this optimization scheme with a neural network ansatz for the time dependent committor allows us to saturate the variational rate bound, and gives us a complete description of the transition path ensemble. Specifically, we have described how to decompose the rate in terms of additive contributions from different degrees of freedom. This  procedure of quantifying contributions can be applied to collective coordinates and order parameters that are used for characterizing reactions of complex molecular systems. We showcase this decomposition by investigating the reaction of Brownian particles in simple potentials in underdamped and overdamped regimes. We have shown how to apply this procedure to conformational changes in solution, leveraging insightful information about the reactive event even when the full phase space is not provided as training data.  In cases where the variational bound is not saturated, the rate can still be computed using other estimators. This decomposition could lend insight into the design of models to accurately recover kinetic information.\cite{bolhuis2022force} 

The formalism employed, casts the time dependent committor as an optimal control force naturally making this model generative. Specifically, when the variational bound is saturated, a time dependent control force is produced that generates reactive trajectories in an unbiased manner. While not used as such here, this procedure can be employed to glean higher order statistics of the reactions over and above the rate.\cite{das2022direct} When the variational bound is not saturated the control force can still be applied to generate unbiased transition path statistics through ensemble \revision{reweighting}.\cite{gao2019nonlinear} One could envision an iterative procedure in cases where path sampling is difficult, for example in cases of long diffusive trajectories, where initial control forces are gradually optimized through alternative cycles of training and reactive ensemble generation. 

As the method is based on ensembles of trajectories and path reweighting, there is no formal restriction to equilibrium systems. Indeed, variational path sampling has been initially applied to systems whose dynamics break detailed balance. As such the procedures developed here for NN based function approximations and rate decompositions transfer over directly to rare transitions in nonequilibrium steady-states. However, traditional path sampling techniques that render the generation of a path ensemble simple in equilibrium are not typically as effective away from equilibrium. \revision{ For those systems, one would have to consider using path sampling methods that do not invoke detailed balance.}\cite{huber1996weighted,guttenberg2012steered,valeriani2007computing} The iterative procedure alluded to above is likely a robust means of extending this methodology to study phase transitions in active matter and driven assembly. 

\section*{Acknowledgements}
We would like to thank Dr. Avishek Das and Dr. Jorge Rosa-Ra\'ices for illuminating discussions regarding variational path sampling methods. We would also like to thank Dr. Clay Batton for useful comments regarding the manuscript. This material is based upon work supported by the U.S. Department of Energy, Office of Science, Office of Advanced Scientific Computing Research and Office of Basic Energy Sciences, via the Scientific Discovery through Advanced Computing (SciDAC) program.

\section*{Data Availability}
The data for the plots and the illustration of this method on the 2D potential can be found within the Github repository \hyperlink{https://github.com/ansingh1214/Deep-VPS}{https://github.com/ansingh1214/Deep-VPS}.
\appendix
\section{Saturation of variational bound}
In Section \ref{Sec:V} we detailed how the time dependent committor can be approximated for a formally underdamped system evolving in an overdamped regime. Here we demonstrate that the approximate form of the time dependent committor saturates the variational rate bound up to order $\mathcal{O}(\gamma^{-2})$. Using the approximation $q(\vc{r}^N,\vc{v}^N,t)= q_0(\vc{r}^N,t) + m \vc{v} \cdot \nabla_{\vc{r}} q_0(\vc{r}^N,t)/\gamma + \mathcal{O}(\gamma^{-2})$ we consider the log transform, $Q=\ln q$, which to equivalent order in perturbation theory is
\begin{align}\label{eq:4.4}
 Q(\vc{r}^N,\vc{v}^N,t)  &\approx \ln \left ( q_0 + \frac{m \vc{v}}{\gamma} \cdot \nabla_{\vc{r}} q_0 + \mathcal{O}(\gamma^{-2}) \right )\notag\\
    &= \ln q_0 + \frac{m \vc{v}}{\gamma} \cdot \nabla_{\vc{r}} \ln q_0 + \mathcal{O}(\gamma^{-2}) 
\end{align}
where we will use $Q_0 = \ln q_0$. For an underdamped equation of motion the relative action, $\Delta U_\lambda$ is given by
\begin{align}
\Delta U_{\lambda}[\mathbf{X}] &= -\sum_{i=1} ^N \frac{1}{4\gamma k_{\mathrm{B}}T} \int_0^{t_f} dt \, \left [ \boldsymbol{\lambda}_i^2 \notag \right .\\
&\qquad \left .- 2\boldsymbol{\lambda}_i \cdot \left(m \dot{\vc{v}}_i + \gamma \vc{v}_i - \vc{F}_i(\vc{r}^N) \right) \right ] \notag 
\end{align}
just as in the overdamped case. Substituting the underdamped Doob force,
\begin{equation}
\boldsymbol{\lambda}_i^*=\frac{2\gamma \kbt}{m} \nabla_{\vc{v}_i} Q(\vc{r}^N,\vc{v}^N,t)
\end{equation}
into $\Delta U_{\lambda}[\mathbf{X}] $ we find,
\begin{align}\label{eq:4.6}
\Delta U_{\lambda^*} &=\sum_i^N \int_0^{t_f} dt \left [ \dot{\vc{v}}_i \cdot \nabla_{\vc{v}_i} Q + \frac{\gamma}{m}{\vc{v}}_i \cdot \nabla_{\vc{v}_i} Q \right .\\
&\qquad \left .- \frac{\vc{F}_i}{m} \cdot \nabla_{\vc{v}_i} Q -\frac{\gamma \kbt}{m^2} (\nabla_{\vc{v}_i} Q)^2 \right ] \notag
\end{align}
The first term can be resolved using Ito's Lemma
\begin{equation}\label{4.7}
\dot Q = \partial_t Q + \sum_i^N \dot{\vc{v}}_i \cdot \nabla_{\vc{v}_i} Q + \vc{v}_i \cdot \nabla_{\vc{r}_i} Q + \frac{\gamma \kbt }{m^2} \nabla_{\vc{v}_i}^2 Q
\end{equation}
Substituting this back to the relative action, we get:
\begin{align}\label{eq:4.8}
    \Delta U_{\lambda^*} &= - \int_0^{t_f} dt \sum_i^N \Big{[}\frac{\gamma \kbt}{m^2} (\nabla_{\vc{v}_i} Q)^2 + \vc{v}_i \cdot \nabla_{\vc{r}_i} Q   \\
    & + \frac{\gamma\kbt}{m^2} \nabla_{\vc{v}_i}^2 Q - \frac{\gamma}{m} {\vc{v}}_i \cdot \nabla_{\vc{v}_i} Q + \frac{\vc{F}_i}{m} \cdot \nabla_{\vc{v}_i} Q\Big{]} \notag \\  &- \dot Q + \partial_t Q \notag
\end{align}
Finally, using the perturbative approximation of $Q$ in Eq. \ref{eq:4.4}, and substituting the approximated form of the backward Kolmogorov equation in Eq. \ref{eq:4.5} yields
\begin{align}\label{4.9}
    \Delta U_{\lambda^*} = -\int_0^{t_f} dt &\sum_i^N \Big{[} \frac{\kbt}{\gamma} (\nabla_{\vc{r}_i} Q_0)^2  + \frac{m \vc{v}_i^2}{\gamma} \nabla_{\vc{r}_i}^2 Q_0 \notag \\
     &+ \frac{\vc{F}_i}{\gamma} \nabla_{\vc{r}_i} Q_0\Big{]} - \dot Q + \partial_t Q  \notag \\
    = \int_0^{t_f}dt \,& \dot{Q} = -\ln q
\end{align}
Hence, $\Delta U_{\lambda^*}$ quantifies the transition probability between the states $A$ and $B$ over time $t_f$ when averaged over an initial distribution in $A$. 

\section{Relative action for Langevin leap-frog integrator}
The equations of motion for the Langevin leap-frog integrator is given by\cite{izaguirre2010multiscale,eastman2017openmm}
\begin{align}
    \mathbf{v}_i[t+\Delta t /2] &= \alpha \mathbf{v}_i[t-\Delta t /2] + \frac{1 - \alpha}{\gamma m_i}\mathbf{F}_i[t] + \boldsymbol{\eta}_i[t] \\ \notag
    \mathbf{r}_i[t + \Delta t] &= \mathbf{r}_i[t] + \mathbf{v}_i[t+\Delta t /2] \Delta t 
\end{align}
where the definitions of $\mathbf{v}_i$, $\mathbf{r}_i$ and $\mathbf{F}_i$ are the same as in Eq. \ref{eq:4.1}, $m_i$ is the mass of particle $i$, $\gamma$ is friction coefficient, $\Delta t$ is the timestep, and $\alpha = \exp[-\gamma \Delta t]$. The noise, $\boldsymbol{\eta_i}$ is a Gaussian random variable with mean $\langle \boldsymbol{\eta}_i(t) \rangle = 0$ and variance $\langle \boldsymbol{\eta}_i(t) \otimes \boldsymbol{\eta}_j(t') \rangle = \kbt (1-\alpha^2)m_i^{-1} \delta_{ij} \mathbf{1} \delta(t-t')$. For this discretization, the relative stochastic action is
\begin{align}
    \Delta U_\lambda =  \sum_n^{t_f/\Delta t} \sum_{i}^N &\frac{  (1-\alpha) \boldsymbol{\lambda_i}^2[n\Delta t] } {2(1+\alpha)m_i\gamma^2k_{\mathrm{B}}T } -  \frac{\boldsymbol{\lambda_i}[n\Delta t] \boldsymbol{\eta_i}[n \Delta t] }{\gamma k_{\mathrm{B}}T (1+\alpha)}
\end{align}
which is the same general form as in the overdamped case.

\section{Internal coordinates for alanine dipeptide}
\begin{figure}
  \centering
    \includegraphics[width=5.5cm]{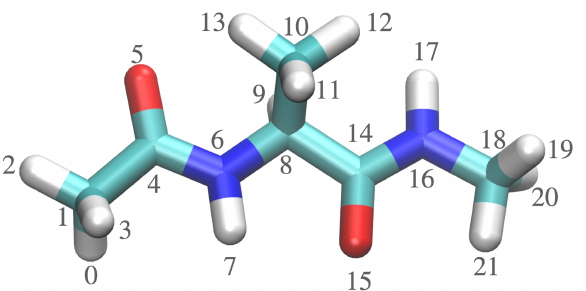}
    \caption{ Atom indices for alanine peptide.}
    \label{fig:9}
\end{figure}

In the studies on alanine dipeptide we parameterized our NN ansatz for the time dependent committor based on a set of internal coordinates. In Tables I and II we define each of the angles and dihedrals referred to in the main text based on the atom numbering Fig.~\ref{fig:9}. 

\setlength{\tabcolsep}{0.5em}
\begin{table}[H] 
\centering
\begin{tabular}{|c|c|c|c|} 
\hline 
Label & Type & Description & Contribution \\
\hline
$d_{4} $ & Dihedral & 4 - 6 - 8 - 14 & 0.416 \\
$d_{3} $ & Dihedral & 4 - 6 - 8 - 10 & 0.160 \\
$d_{8} $ & Dihedral & 10 - 8 - 14 - 16 & 0.079 \\
$a_{3} $ & Angle & 5 - 4 - 6 & 0.048 \\
$d_{11} $ & Dihedral & 1 - 6 - 4 - 5 & 0.043 \\
$d_{7} $ & Dihedral & 10 - 8 - 14 - 15 & 0.042 \\
$a_{1} $ & Angle & 1 - 4 - 5 & 0.025 \\
$a_{8} $ & Angle & 8 - 14 - 15 & 0.023 \\
$a_{10} $ & Angle & 15 - 14 - 16 & 0.022 \\
$a_{11} $ & Angle & 14 - 16 - 18 & 0.021 \\
$a_{9} $ & Angle & 8 - 14 - 16 & 0.021 \\
$a_{6} $ & Angle & 6 - 8 - 14 & 0.019 \\
$a_{2} $ & Angle & 1 - 4 - 6 & 0.018 \\
$d_{12} $ & Dihedral & 8 - 16 - 14 - 15 & 0.014 \\
$a_{4} $ & Angle & 4 - 6 - 8 & 0.011 \\
$a_{7} $ & Angle & 10 - 8 - 14 & 0.009 \\
$d_{9} $ & Dihedral & 8 - 14 - 16 - 18 & 0.004 \\
$d_{10} $ & Dihedral & 15 - 14 - 16 - 18 & 0.002 \\
$a_{5} $ & Angle & 6 - 8 - 10 & 0.002 \\
$d_{2} $ & Dihedral & 5 - 4 - 6 - 8 & 0.001 \\
$d_{1} $ & Dihedral & 1 - 4 - 6 - 8 & -0.009 \\
$d_{6} $ & Dihedral & 6 - 8 - 14 - 16 & -0.010 \\
$d_{5} $ & Dihedral & 6 - 8 - 14 - 15 & -0.012 \\
\hline 
\end{tabular} 
\caption{Contribution to rate in implicit solvent}\label{Tab1}
\end{table}

\begin{table}[H] 
\centering
\begin{tabular}{|c|c|c|c|} 
\hline
Label & Type & Description & Contribution\\
\hline
$a_{13} $ & Angle & 0 - 1 - 3 & 0.035 \\
$a_{41} $ & Angle & 17 - 16 - 18 & 0.035 \\
$d_{4} $ & Dihedral & 4 - 6 - 8 - 14 & 0.034 \\
$a_{23} $ & Angle & 7 - 6 - 8 & 0.032 \\
$a_{33} $ & Angle & 11 - 10 - 12 & 0.032 \\
$a_{12} $ & Angle & 0 - 1 - 2 & 0.031 \\
$a_{15} $ & Angle & 2 - 1 - 3 & 0.030 \\
$a_{34} $ & Angle & 11 - 10 - 13 & 0.029 \\
$d_{21} $ & Dihedral & 5 - 4 - 6 - 7 & 0.026 \\
$a_{47} $ & Angle & 20 - 18 - 21 & 0.025 \\
$d_{7} $ & Dihedral & 10 - 8 - 14 - 15 & 0.024 \\
$d_{3} $ & Dihedral & 4 - 6 - 8 - 10 & 0.023 \\
$d_{11} $ & Dihedral & 1 - 6 - 4 - 5 & 0.023 \\
$a_{27} $ & Angle & 9 - 8 - 10 & 0.021 \\
$d_{23} $ & Dihedral & 4 - 6 - 8 - 9 & 0.019 \\
$a_{35} $ & Angle & 12 - 10 - 13 & 0.018 \\
$a_{44} $ & Angle & 16 - 18 - 21 & 0.017 \\
$d_{33} $ & Dihedral & 9 - 8 - 10 - 12 & 0.015 \\
$a_{42} $ & Angle & 16 - 18 - 19 & 0.015 \\
$d_{17} $ & Dihedral & 3 - 1 - 4 - 5 & 0.015 \\
\hline
\end{tabular} 
\caption{Contribution to rate in explicit solvent}\label{Tab2}
\end{table}

\bibliography{DeepVPS}

\end{document}